\documentclass[apj]{emulateapj}

\newcommand {\Lya}    {Ly$\alpha$}   
\newcommand {\Lyb}    {Ly$\beta$}    
\newcommand {\Lyg}    {Ly$\gamma$}

\newcommand {\Lye}    {Ly$\epsilon$}

\newcommand {\HI}     {\ion{H}{1}}   
\newcommand {\HII}    {\ion{H}{2}}   
\newcommand {\OI}     {\ion{O}{1}}   
\newcommand {\OVI}    {\ion{O}{6}}   
\newcommand {\OVII}   {\ion{O}{7}}
\newcommand {\OVIII}  {\ion{O}{8}}
\newcommand {\CIII}   {\ion{C}{3}}   
\newcommand {\NV}     {\ion{N}{5}}
\newcommand {\CIV}    {\ion{C}{4}}
\newcommand {\SiIV}   {\ion{Si}{4}}
\newcommand {\SiIII}  {\ion{Si}{3}}
\newcommand {\SiII}   {\ion{Si}{2}}

\newcommand {\NeVIII} {\ion{Ne}{8}}
\newcommand {\SIII}   {\ion{S}{3}}
\newcommand {\SIV}    {\ion{S}{4}}

\newcommand {\Wlya}   {W_{\rm Ly\alpha}} 
\newcommand {\Wlyb}   {W_{\rm Ly\beta}} 
\newcommand {\Wlyg}   {W_{\rm Ly\gamma}} 
\newcommand {\blyb}   {b_{\rm Ly\beta}} 
\newcommand {\kms}    {km~s$^{-1}$}
\newcommand {\NHI}    {$N_{\rm HI}$}
\newcommand {\NOVI}   {$N_{\rm OVI}$}

\newcommand {\bHI}    {$b_{\rm HI}$}
\newcommand {\bCOG}   {b_{\rm COG}} 
\newcommand {\btherm} {b_{\rm T}} 
\newcommand {\bLW}    {b_{\rm LW}} 
\newcommand {\bLya}   {b_{\rm Ly\alpha}} 
\newcommand {\tnma}{\tablenotemark{a}}
\newcommand {\tnmb}{\tablenotemark{b}}
\newcommand {\tnmc}{\tablenotemark{c}}

\newcommand {\lam}    {$\lambda$}
\newcommand {\FUSE}  {{\it FUSE}}

\newcommand {\dndz}  {$d{\cal N}/dz$}

\newcommand {\etal}  {et~al.} 
\newcommand {\cd}    {cm$^{-2}$}  

\slugcomment{ApJ, 1st Revision}

\begin{document}

\title{Broad H\,I Absorbers as Metallicity-Independent Tracers of the Warm-Hot Intergalactic Medium\footnote{Based on observations made with the NASA/ESA Hubble Space Telescope, obtained at the Space Telescope Science Institute, which is operated by the Association of Universities for Research in Astronomy, Inc., under NASA contract NAS 5-26555.}}
\author{Charles W. Danforth, John T. Stocke, \& J. Michael Shull}
\affil{CASA, Department of Astrophysical and Planetary Sciences, University of Colorado, 389-UCB, Boulder, CO 80309; danforth@casa.colorado.edu, john.stocke@colorado.edu, michael.shull@colorado.edu}


\begin{abstract}
Thermally broadened \Lya\ absorbers (BLAs) offer an alternative method to highly-ionized metal lines for tracing the warm-hot intergalactic medium (WHIM) at $T>10^5$~K.  However, observing BLAs requires data of high quality and accurate continuum definition to detect the low-contrast features, and a good knowledge of the velocity structure to differentiate multiple blended components from a single broad line.  Even for well-characterized absorption profiles, disentangling the thermal line width from the various thermal and non-thermal contributors to the observed line width is ambiguous.  We compile a catalog of reliable BLA candidates along seven AGN sight lines from a larger set of \Lya\ absorbers observed by the Space Telescope Imaging Spectrograph (STIS) on the Hubble Space Telescope (HST).  We compare our measurements based on independent reduction and analysis of the data to those published by other research groups.  We examine the detailed structure of each absorber and determine a reliable line width and column density.  Purported BLAs are grouped into probable (15), possible (48) and non-BLA (56) categories.  Combining the first two categories, we infer a line frequency $(d{\cal N}/dz)_{\rm BLA}=18\pm11$, comparable to observed \OVI\ absorbers, also thought to trace the WHIM.  We discuss the overlap between BLA and \OVI\ absorbers (20-40\%) and the distribution of BLAs in relation to nearby galaxies (\OVI\ detections in BLAs are found clouser to galaxies than \OVI\ non-detections).  We assume that the line width determined through a multi-line curve-of-growth (COG) is a close approximation to the thermal line width.  Based on 164 measured COG \HI\ line measurements, we statistically correct the observed line widths via a Monte-Carlo simulation.  Gas temperature and neutral fraction $f_{\rm HI}$ are inferred from these statistically-corrected line widths and lead to a distribution of total hydrogen columns.  Summing the total column density over the total observed pathlength, we find a BLA contribution to the closure density of $\Omega_{\rm BLA}=6.3^{+1.1}_{-0.8}\times10^{-3}\,h_{70}^{-1}$ based on $10^4$ Monte-Carlo simulations of each BLA system.  There are a number of critical systematic assumptions implicit in this calculation, and we discuss how each affects our results and those of previously published work.  In particular, the most comparable previous study by \citet{Lehner07} gave $\Omega_{\rm BLA}=3.6\times10^{-3}\,h_{70}^{-1}$ or $9.1\times10^{-3}\,h_{70}^{-1}$, depending on which assumptions were made about hydrogen neutral fraction.  Taking our value, current \OVI\ and BLA surveys can account for $\sim20$\% of the baryons in the local universe while an additional $\sim29$\% can be accounted for in the photoionized \Lya\ forest; about half of all baryons in the low-$z$ universe are found in the IGM.  Finally, we present new, high-S/N observations of several of the BLA candidate lines from Early Release Observations made by the Cosmic Origins Spectrograph on HST.
\end{abstract}



\section{Introduction}

Theoretical studies of cosmological structure formation and the intergalactic medium (IGM) predict that the baryonic component should develop a hot, shocked phase at temperatures of $T\ga10^5$~K.  Adding to these gravitation shocks is the energy feedback from galactic winds, which can produce a hot circumgalactic medium extending 100-200 kpc from galaxies.  Understanding this warm-hot intergalactic medium (WHIM) is therefore crucial to understanding local large-scale structure, galaxy feedback, and the spread of metals.  Because the IGM in the early universe ($z=2-6$) was primarily warm and photoionized, \Lya\ forest absorption can be used to trace the large-scale structures at those times accounting for the majority of the baryons \citep{Rauch98}.  However, the fraction of baryons in the \Lya\ forest is observed to change in form and content toward lower redshifts owing to the rapid drop of the extragalactic ionizing background and the formation of large-scale structure \citep{Penton04,Weymann01}.  Cosmological simulations \citep[e.g.,][]{Dave99,Dave01,CenOstriker99,CenFang06} predict that by $z\sim0$ the baryons are distributed among three dominant reservoirs: 10--30\% in cool gas ($\leq 10^4$ K) in and around galaxies, 30\% remaining in the warm, photoionized \Lya\ absorbing gas, and the remainder in hotter WHIM gas.  These same simulations predict that shocks from diffuse matter falling onto large-scale structure filaments and from galactic wind feedback into these filaments have heated a large fraction of the low-$z$ baryons.  While a broad spectrum of shock velocities undoubtedly produces some shocked material at cooler temperatures, the WHIM phase is defined as low-density gas at $T=10^5-10^7$~K.  Two recent simulations estimate the baryon fraction residing in the WHIM at $30-50$\%, but with considerable uncertainty \citep{Dave01,CenFang06} largely due to the pervasiveness and effectiveness of feedback from galaxies into the IGM.  Thus, no accurate baryon census can be made without detailed knowledge of WHIM absorbers..

Three observational methods have been employed to detect and obtain a rough census of the WHIM.  Ions with high ionization potentials ($\ga100$ eV) are likely the product of collisional ionization and thus trace hot gas.  Absorption of highly-ionized metals can in principle be observed in the X-ray; in particular \OVII\ (21.60~\AA) and \OVIII\ (18.97~\AA) since they are strong transitions of an abundant element with peak collisional ionization equilibrium (CIE) abundances at temperatures of $(1-2)\times10^6$~K.  Cooling of diffuse gas at these temperatures is slow and a significant gas reservoir is expected.  While there have been several reported intergalactic \OVII\ and \OVIII\ detections \citep{Fang02,Fang07,Nicastro05} based on {\it Chandra} observations along AGN sight lines, none has been confirmed with {\it XMM/Newton} spectra \citep{Rasmussen07,Kaastra06} and their reality has been questioned \citep{Bregman07,Richter08}.  \citet{Yao09} used a statistical ``stack-and-add'' technique looking for \OVII\ absorption associated with known \OVI\ systems to place upper limits of $N_{\rm OVII}\la 10\times N_{\rm OVI}$.  However, definitive X-ray WHIM detections have remained, at best, controversial.

Less highly-ionized species such as \OVI\ provide the second WHIM tracer.  Until now, this has proven to be the most fruitful technique due to extensive use of the UV spectrographs on board the {\it Hubble Space Telescope (HST)} and the {\it Far Ultraviolet Spectroscopic Explorer (FUSE)}.  Absorption in the Li-like doublets \OVI\ \lam\lam1032,1038 and \NV\ \lam\lam1238,1242 traces gas with peak CIE abundances at the lower end of the WHIM temperature range ($10^{5-6}$~K).  Several large \OVI\ absorption-line surveys of the local IGM toward bright AGN have been published.  \citet[hereafter DS08]{DS08} report 83 \OVI\ absorption systems out of a total of $\sim$650 \Lya\ absorbers along AGN sight lines surveyed at $z<0.4$.  \citet{Tripp08} and \citet{ThomChen08a} used somewhat smaller numbers of sight lines, but report 51 and 27 \OVI\ absorbers, respectively, with detection statistics similar to DS08.  Two \ion{Ne}{8} \lam\lam 760, 770 detections ($T_{\rm max}\sim10^{5.85}$~K) have been made \citep{Savage05,Narayanan09}.  These may provide a more reliable WHIM tracer, but the current statistics are poor.

There are several controversies, weaknesses, and unknowns that underlie UV metal-line WHIM surveys \citep[see ][]{Danforth09}.  While the number of absorbers is generally agreed upon, their interpretation remains controversial;  \OVI, \NV, and \CIV\ ions are produced readily in collisionally ionized gas, but they can also be photoionized by sufficiently energetic photons (114, 78, and 48 eV, respectively).  The observed hydrogen and oxygen column densities require very diffuse ($n_H<10^{-5}\rm~cm^{-3}$) gas over long pathlength regions ($\ga 500\,h^{-1}_{70}$~kpc) to reproduce the line strengths and ratios seen in \Lya\ and \OVI\ absorbers \citep[e.g.,][]{Prochaska04,Tripp08,OppenheimerDave08}.  Thus, some high-ion absorbers may not be shock-heated WHIM material, but instead low density, photoionized gas.  

Regardless of the ionization state, it is important to note that \OVI\ absorbers are found in locations relatively near bright ($L^*$) galaxies \citep[$\leq800\,h^{-1}_{70}$~kpc][]{Stocke06,WakkerSavage09} as expected for metal-enriched WHIM gas based upon numerical simulations \citep{Dave99, CenOstriker06, CenFang06}.  Since some degree of metal enrichment is required for this WHIM diagnostic (the ``metallicity bias''), we cannot use this method to sample the primordial IGM far from metal production sites in galaxies and beyond the range of metal distribution via starburst winds \citep{Stocke06,Stocke07}.  Thus, this method cannot possibly sample all WHIM gas.  Finally, translating metal-ion statistics into a total baryon distribution requires estimating the metallicity $Z$ and fractional ion abundance $f_{\rm ion}$, entailing additional assumptions.  Even the best statistics for \OVI\ line densities (\dndz) may result in a factor-of-two uncertainty in the final baryon budget.

An orthogonal route to detecting WHIM absorbers, and one that does not rely upon metal enrichment, is to detect and measure broad \Lya\ absorbers (BLAs) in high-S/N spectra.  Hydrogen is so abundant in the universe that, even with a neutral fraction of $10^{-5}$ or less, there can be measurable absorption in \Lya\ lines.  There is no ``metallicity bias'' and no abundance uncertainty.  In collisionally ionized gas, the neutral fraction is a function of temperature and can, in principle, be directly measured from the line width.  Line width is usually denoted in terms of the corresponding doppler parameter $b\equiv\rm FWHM/2\sqrt{\ln2}\equiv\sqrt{2}\,\sigma$ where $\sigma$ is the Gaussian width.  Thermal width is simply a function of gas temperature $T$ and atomic mass $A$; 
\begin{eqnarray}
  b_{\rm T}(T)&=&\sqrt{2kT/m}=\sqrt{T/60\,A}~\rm km~s^{-1}\nonumber
\\ &=&(40.6~{\rm km~s^{-1}})\,T_5^{1/2}\,A^{-1/2},\label{eq_bt}
\end{eqnarray}
where $T_5$ is temperature in units of $10^5$~K. 

HST has observed over 100 bright AGN sight lines at modest signal-to-noise (S/N) ratios ($5-20$) with its three generations of spectrometers: the Faint Object Spectrograph (FOS), the Goddard High Resolution Spectrograph (GHRS), and the Space Telescope Imaging Spectrograph (STIS).  Until the recent installation of the Cosmic Origins Spectrograph, the best data for a BLA search comes from recent STIS echelle spectroscopy, which typically features the highest S/N combined with a spectral resolution sufficient to resolve \HI\ lines of widths less than $\sim30$~\kms.  From detailed analyses of moderate S/N STIS echelle spectra of individual bright AGN, \citet{Sembach04} and \citet{Tripp01} noted the presence of a few very broad \Lya\ lines confirmed to have $b>40$ \kms\ using curve-of-growth (COG) analyses of multiple Lyman lines.  \citet{Richter04}, \citet{Aracil06a}, \citet{Lehner06}, and \citet{Williger06} presented BLA candidates based upon line-width measurements, as did \citet{Penton04}.  \citet{Richter06a} compiled a few of the very best cases of BLAs.  More recently \citet[henceforth L07]{Lehner07} produced a large compilation of 7 AGN sight lines and $\sim100$ candidate BLAs based on $b$-values $\geq40$ \kms\ (defined variously; see below). 

However elegant, BLA surveys are not without their practical complications, as we describe here and in Section~2.1.  First, identifying {\it bona fide}, thermally-broadened \HI\ systems is challenging.  Absorption features arising from trace neutral hydrogen in WHIM should exhibit broad profiles and low column densities, which are difficult to detect in spectra of finite S/N.  Additionally, differentiating single, broad features from blended profiles or those with ambiguous component structure is central to their identification as BLAs.  Because of these ambiguities, we provide an independent opinion on many previously studied sight lines.  Second, inferring a temperature from a measured line width is crucial for inferring the total gas column density present.  Most previous authors recognize that the observed line width $\bLW$ is, at best, an upper limit on the thermal line width, so that only upper limits on total baryon mass can be inferred.  For stronger \Lya\ absorbers, multiple Lyman lines allow us to measure the curve of growth (COG) for cool \HI\ systems and then apply a statistical correction to obtain better estimates on the distribution of $\btherm$ (Section 3.3).  Finally, determining a hydrogen neutral fraction $f_{\rm HI}$ from an inferred temperature depends on assumptions about collisional (thermal) ionization and photoionization by the metagalactic radiation field.  In Section 4.1, we address the relative importance of these two mechanisms from first principles.

Owing to the difficulties in definitively detecting BLAs and the clear importance in doing so, we take a detailed look at broad \HI\ systems from our own work and other published studies.  In Section~2 we discuss our methodology, including a discussion of the complications surrounding BLA surveys.  Our results are presented in Section 3, including a cross-correlation of our BLA catalog with recent large surveys of low-$z$ IGM metal line absorbers (DS08) and nearby galaxies.  We discuss the larger cosmological conclusions of this BLA census and its systematic uncertainties in Section~4.  Our conclusions are presented in Section~5.   Those interested in an overview of our results are cautioned against skipping directly to Section~3, since understanding our selection criteria for BLA vs. non-BLA absorbers is critical to understanding our results in general.

\section{Methodology and Absorber Sample}

\subsection{Broad \Lya\ Lines as a WHIM Tracer} 

\begin{figure}
  \epsscale{1.2}\plotone{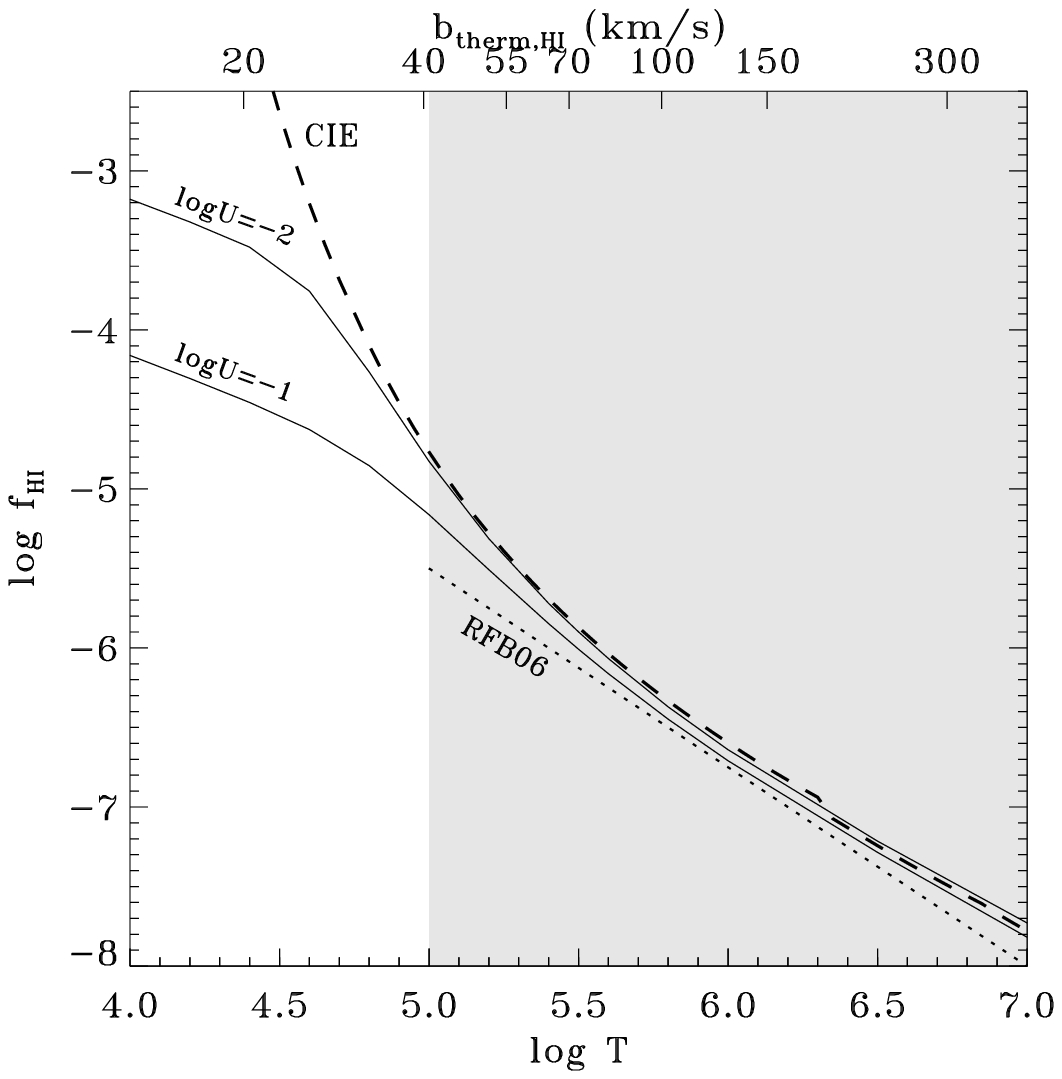} 
  \caption{The neutral fraction $f_{\rm HI}$ of hydrogen drops sharply
  as a function of temperature.  In collisional ionization equilibrium
  \citep[dashed line,][]{SutherlandDopita93} $f_{\rm HI}\la10^{-5}$ in
  the WHIM temperature range ($T=10^5-10^7$~K, shaded region).
  \citet{Richter06b} parametrize CIE+photoionization simulation
  results to a linear relationship with $\log\,T$ (dotted line) that
  is $\sim2-6$ times smaller than CIE at $T=(1-3)\times10^5$~K.
  CLOUDY photo$+$thermal ionization models \citep{Paper2} with
  photoionization parameters $\log\,U=-2$ and $-1$ (solid lines)
  approximate the simple CIE curve throughout the WHIM range (shaded),
  but deviate significantly at lower temperatures.  Thermal $b$-value
  for \HI\ is shown on the top axis.}  \label{fig_fhi}
\end{figure}

Using thermally-broadened \HI\ lines to trace WHIM gas avoids many of the biases that can plague metal-ion based WHIM tracers \citep[see][]{Danforth09}.  However, BLA surveys are fraught with observational complications as well \citep[e.g.][]{Richter04,Richter06a}.  First, BLA lines are hard to detect and require spectra with a high signal-to-noise ratio (S/N) and minimal instrumental systematics.  For example, at $T=10^5$~K, the hydrogen neutral fraction is approximately $f_{\rm HI}\sim10^{-5}$ (Fig.~\ref{fig_fhi}) and the thermal $b$-value is 40~\kms\ (Eq.~\ref{eq_bt}).  An IGM absorber with a total hydrogen (\HI$+$\HII) column density $N_{\rm H}=10^{18}$~\cd, typical of what is expected in the \Lya\ forest, would still have an observable \HI\ column of $N_{\rm HI}\sim10^{13}$~\cd.  In the optically thin regime (linear COG of growth), the equivalent width $W_\lambda$ and line-center optical depth $\tau_0(N,T)$ for a Gaussian profile BLA are
\begin{equation}
  W_\lambda(N)=\Bigl(\frac{\pi e^2}{m_e c}\Bigr)\,\Bigl(\frac{N f \lambda^2}{c}\Bigr)=(54.5~{\rm m\AA})\,N_{13} \label{eq_wbla}
\end{equation}
\begin{equation}
  \tau_0(N,T)=\frac{\pi e^2}{m_e c} \frac{N f \lambda^2}{\sqrt{\pi} b}=(0.187)\,N_{13}\,T_5^{-1/2},\label{eq_taubla}
\end{equation}
where $N_{13}$ is the \HI\ column density in units of $10^{13}$~\cd.  Thus, an IGM absorber with $N_{13}\approx1$ will have a rest-frame \Lya\ equivalent width $W_\lambda=55$~m\AA\ and a fractional depth of $\sim20$\% (assuming no non-thermal broadening).  This line would be easily observable in data of modest S/N.  

However the prospects get much worse at higher temperatures, as the thermal line width rises as $\sqrt{T}$ and the neutral fraction drops (Fig.~\ref{fig_fhi}).  A \Lya\ absorber with the same total hydrogen column as above but with a temperature of $3\times10^5$~K (near the peak CIE abundance of \OVI) would have $f_{\rm HI}\sim10^{-6}$, $N_{\rm HI}\sim10^{12}$~\cd, $b=70$~\kms, $W_\lambda\sim5$~m\AA, and a fractional depth $\tau_0\sim1$\%, impossible to detect unless the spectra have $S/N\gg100$.  Turbulent and/or bulk cloud motions would broaden the \Lya\ line even further and reduce the fractional line depth.  Confirmation of a \Lya\ absorber and definitive measurements of \bHI\ and \NHI\ using higher Lyman lines and COG techniques are difficult for such lines since even \Lyb\ is a factor $\sim7$ weaker than \Lya\ for unsaturated systems.

\begin{figure*} 
\epsscale{1}\plotone{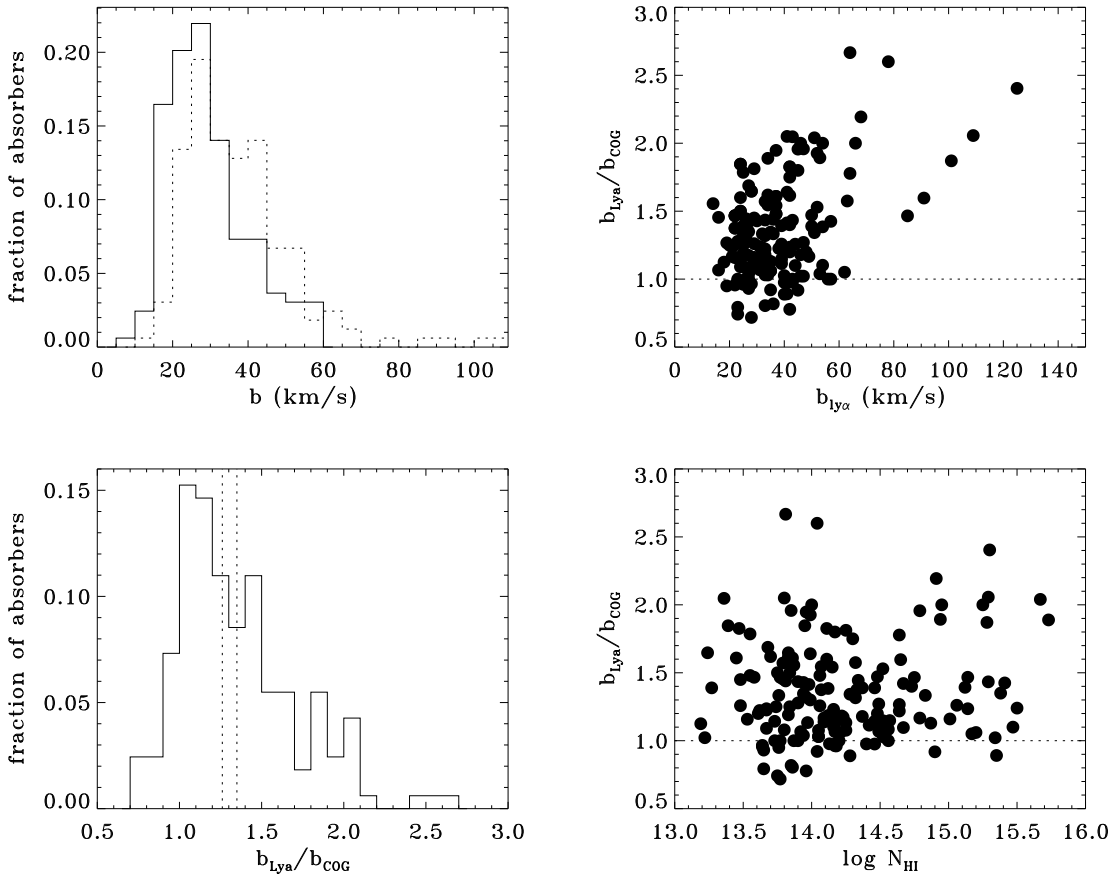} 
  \caption{Comparison of line width $\bLW$ and COG doppler parameter
  $\bCOG$ in 164 \Lya\ absorbers from DS08.  Top left: histogram of
  measured $\bLW$ (dotted line) and $\bCOG$ (solid line).  Bottom
  left: ratio $\bLW/\bCOG$ covers a wide range, with a mean 1.35
  and median $1.26^{+0.49}_{-0.25}$ (dotted lines).  The top right
  panel shows that $\bLW/\bCOG$ is weakly correlated with $\bLW$
  ($R=0.45$).  Lower right panel shows that $\bLW/\bCOG$ is
  uncorrelated with $\log\,N_{\rm HI}$ ($R=0.1$).}\label{fig_DS08}
\end{figure*}

Detecting low-contrast absorption features depends critically on obtaining a correct model of the continuum level, as well as a presumption that the intrinsic continuum emanating from the AGN central engine is flat and featureless over these broad line widths.  Since the UV continuum of Seyferts and QSOs is thought to arise in thermal accretion disk emission close to the central black hole, this requires that the accretion disk itself be featureless, which may or may not be a correct assumption.  For this reason, the UV continua of BL\,Lac objects may provide better background sources for sensitive BLA detection.  Additionally, fixed pattern noise and echelle-blazed gratings can themselves produce gentle undulations in the continuum, either on the detector or in the extraction process of curved orders imaged on rectilinear detectors and their corresponding artifacts and blemishes.  The possible inability to place an accurate continuum due to the above difficulties not only makes low-contrast line detection suspect, but also makes it hard to set sensible detection limits as a function of $b$-value.

Even assuming data with sufficiently high S/N and a well-defined continuum, the identification of BLAs relies crucially on deconvolving the thermal line width $b_{\rm T}$ from the total observed line width $b_{\rm obs}$.  A simple, single-component absorber can be modeled as a Voigt profile, and the observed line width is the quadrature sum of the thermal line width $\btherm$, the instrumental point spread function (usually approximated as a Gaussian $b_{\rm inst}$), and non-thermal broadening term $b_{\rm NT}$.  The latter term encompasses bulk turbulence, multiple unresolved velocity components, or any other condition that will broaden a line profile.  Instrumental broadening is generally well determined and can be subtracted from the total line width.  However, the relative proportion of the thermal and non-thermal contributions is typically degenerate for absorption in a single species.  

Some simulations \citep[e.g.,][]{Richter06b} suggest that turbulence contributes on average $\sim10$\% to the total line width in broad absorbers, while other simulations suggest that it may be as much as 50\% \citep{CenFang06}.  Observational studies have shown that the line width measured from a single \Lya\ line is generally greater than that determined through a multi-Lyman-series COG \citep{Shull00}.  For example, the so-called Virgo Cluster absorbers in the high S/N GHRS spectrum of 3C\,273 show a pair of \Lya\ components with measured $b=41$~\kms\ and $34$~\kms\ for the 1015~\kms\ and 1590~\kms\ IGM absorbers, respectively \citep{Weymann95}.  However, subsequent {\it FUSE} observations \citep{Sembach01} determined $\bCOG=30$~\kms\ and 16~\kms\ using three and eight higher-order Lyman lines, respectively.  

While these absorbers make good individual examples, statistical studies show that $b$-values determined from line-width measurements of single lines, hereafter denoted $\bLW$, systematically overpredict the true line width.  \citet{Shull00} found that $\bLW/\bCOG\sim2$ using \Lya\ and \Lyb\ measurements of 12 absorbers.  \citet{Paper2} confirmed the low-$z$ result with a larger sample ($\sim100$ absorbers).  Similar conclusions were reached by \citet{Songaila97,Songaila98,Songaila01} for high-$z$ \Lya\ forest lines.  We refine this relationship further using the large catalog of DS08.  Out of $\sim650$ \HI\ absorbers in DS08 plus additions, we measured 164 in multiple Lyman lines, and a COG analysis was performed giving more accurate values of \bHI\ and \NHI.  Figure~\ref{fig_DS08} shows the distribution of $\bLya$ and $\bCOG$ for this sample.  Note that there is no clear correlation of $\bLW/\bCOG$ with column density and only a weak trend with $\bLW$.  The median overprediction ratio is $\bLW/\bCOG=1.26^{+0.49}_{-0.25}$ with a mean of 1.35.  We will use this result extensively in our analysis for the BLAs in which only \Lya\ measurements are available.

Another impediment to BLA identification and measurement is that \Lya\ absorption lines often show ambiguous component structure.  Several narrow components in a close blend can artificially broaden a line.  Sometimes this will manifest itself as an asymmetric or obviously non-Gaussian line profile, but this blending can be undetectable even at arbitrarily high S/N and good spectral resolution.  For this reason, even COG-determined $b$-values can overpredict the thermal line width.  

\begin{figure*}
  \plottwo{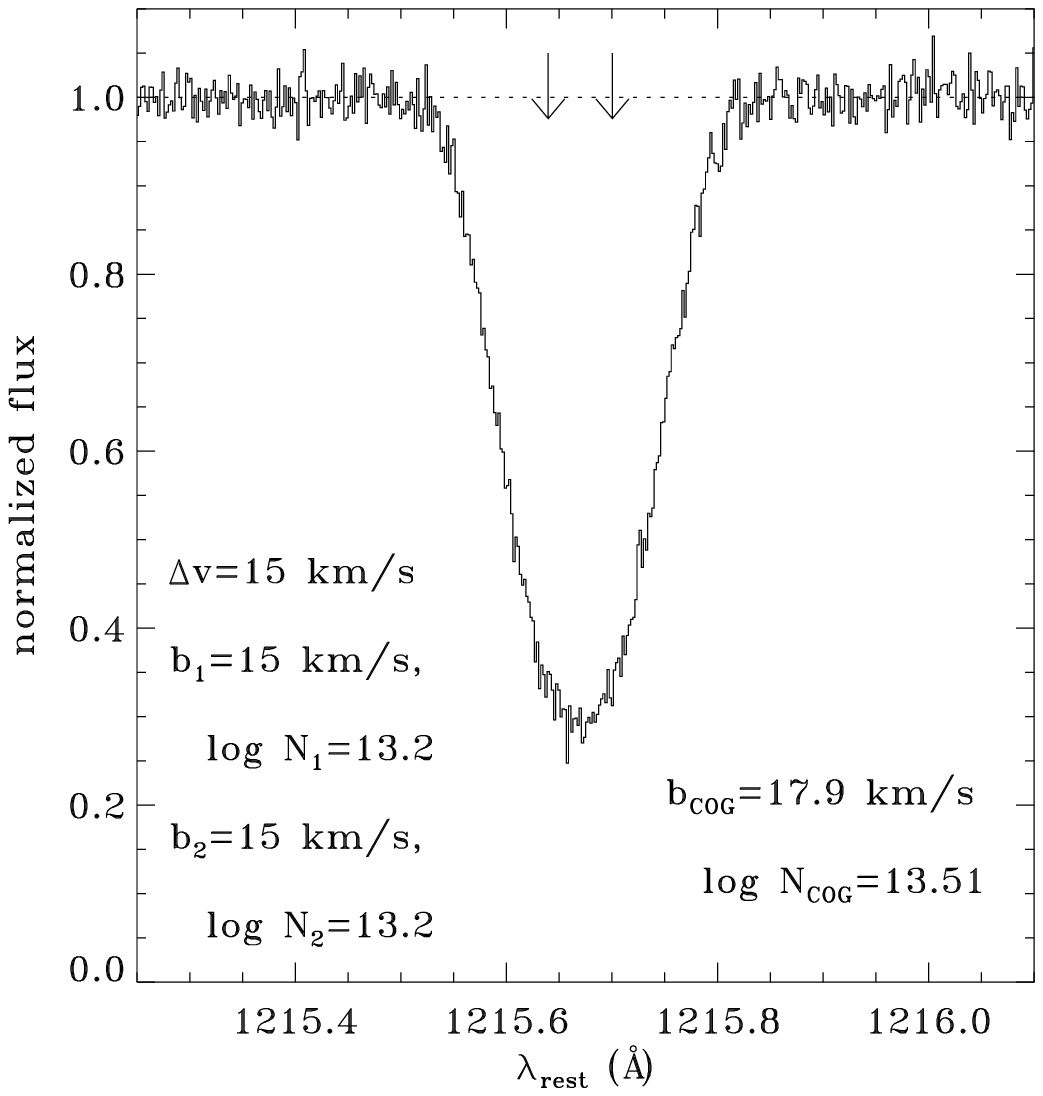}{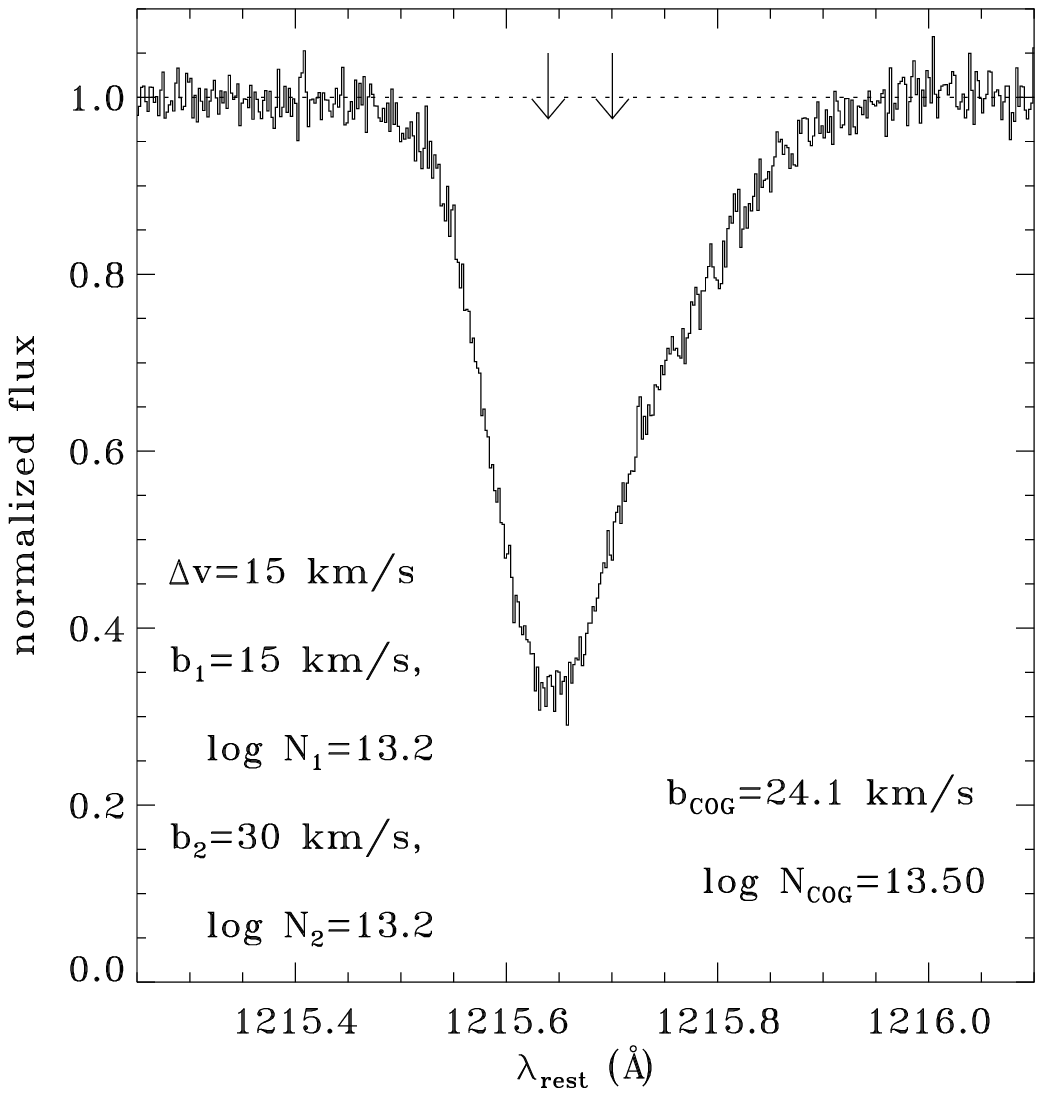} 
  \caption{Blended absorption components in simulated STIS/E140M data.
  Two narrow components (left, $b_1=b_2=15$~\kms) separated by $\Delta
  v=b_1$ (component centroids marked with arrows) produce a combined
  profile which is indistinguishable from a single, broader component
  even in data of high S/N and resolution ($S/N=30$ per 7~\kms\
  resolution element).  Fitting the profile as a single line gives
  $b_{\rm Ly\alpha}=19$~\kms\ or a 27\% overestimate of $b_1$.  Even a
  COG analysis using \Lya\ and \Lyb\ profiles gives $b_{\rm
  COG}\sim1.2\,b_1$.  Doubling the line width of one component (right,
  $b_2=2\,b_1=30$~\kms) simulates a multiphase, broad + narrow system.
  The asymmetric profile makes it obvious that a single-component fit
  is not appropriate, but the component deconvolution and number of
  components are ambiguous.}  \label{fig_blends}
\end{figure*}

To estimate the degree of line-width overprediction from COG analysis, we have modeled two simple two-component systems where the line width and relative centroid separation are varied \citep[see also][]{Paper2}.  In case 1 (Figure~\ref{fig_blends}a), two narrow components, each with thermal line width $b_{\rm T}=15$ \kms, are blended into a single absorption line in \Lya\ and higher Lyman lines.  The equivalent widths of each blended Lyman line are fitted with a COG.  For component separation $\Delta v<b_{\rm T}$, the inferred $\bCOG$ overpredicts $b_{\rm T}$ by $<20\%$, but at $\Delta v>b_{\rm T}$, the overprediction rises to nearly 50\%.  This is largely independent of the relative column densities of the two lines.  Depending on the strength of the line, the resolution of the spectrograph, and the quality of the data, even separations of $\Delta v>b_{\rm T}$ are not obvious in the line profile.  Thus, a pair of narrow lines can easily be mistaken for a single broader system, even using a COG.

In case 2 (Figure~\ref{fig_blends}b), we combine a narrow system ($b_{\rm T,1}=15$ \kms) with a broad absorber ($b_{\rm T,2}=2\,b_{\rm T,1}$) and vary the relative line strengths and component separations.  For components of similar column density, $\bCOG$ is close to the mean of the two components, and separating the components by $(1-2)\,b_{\rm T,1}$ increases the composite $\bCOG$ by $\sim20-50\%$ as in case 1.  When one component is considerably stronger than the other, it dominates the $\bCOG$ solution.  It is reassuring that the total column density of the composite system is conserved, independent of component separation and relative strength \citep{Jenkins86}.  

Appropriate to case 2, many observed absorbers classified as WHIM due to their \OVI\ absorption are multiphase in nature \citep[e.g.,][]{Paper2,DS08,Tripp08}.  A cool, photoionized component ($T<10^5$~K) is present at nearly the same velocity as a warm-hot component.  We now estimate the broad \Lya\ profile expected to correspond with observed WHIM gas.  For a typical \OVI\ absorber with $N_{\rm OVI}\approx(10^{13.5}~{\rm cm^{-2}})\,N^{\rm (OVI)}_{13.5}$ at the peak CIE \OVI\ abundance temperature ($T\approx T_{\rm max}=10^{5.45}$~K) the resulting BLA should have total hydrogen column density
\begin{eqnarray}
  N_{\rm hot}^{\rm (H)}&=&\frac{N_{\rm OVI}}{\langle f_{\rm OVI} \rangle (4.90\times10^{-5})\,\bigl(\frac{Z}{0.1\,Z_\sun}\bigr)}\nonumber \\
&=&(2.93\times10^{18}~{\rm cm^{-2}})\,N_{13.5}^{\rm (OVI)}\,\Bigl(\frac{Z}{0.1\,Z_\sun}\Bigr)^{-1},
\end{eqnarray} 
and, from Eq.~\ref{eq_bt}, $b_{\rm HI}=(68.2~{\rm km~s^{-1}})\,(T/T_{\rm max})^{1/2}$.  Here we adopt an oxygen abundance $(O/H)=(4.90\times10^{-5})(Z/0.1Z_\sun)$ scaled to 10\% of the solar metallicity \citep{Asplund05}.  Given a typical neutral fraction $f_{\rm HI}\approx1.4\times10^{-6}$ at $T\approx T_{\rm max}$, the resulting BLA will have $N_{\rm HI}\sim5\times10^{12}$ \cd\ and $\tau_0\sim0.05$, easily overwhelmed by the narrower, stronger \HI\ absorption from the photoionized component (typically $N_{\rm HI}\approx10^{13-15}$~\cd, $b=20-30$ \kms, $\tau_0\gg1$).  The BLA signature will be apparent only in the line wings and will require exquisite data (S/N $\geq30-40$) and a very good knowledge of the instrumental point spread function to recover.  

Multiphase systems are difficult to identify individually, but there is statistical evidence for broad-plus-narrow \HI\ systems \citep[e.g.,][]{Tripp01,DS08,Tripp08}.  DS08 compared the \bHI\ distribution of 83 \HI\ systems with \OVI\ detections with another 273 having clean \OVI\ nondetections ($N_{\rm OVI}<10^{13.2}$~\cd).  For the \OVI\ detections, the median and standard deviation were $b_{\rm HI}=31\pm15$~\kms, while the \OVI\ nondetections show $b_{\rm HI}=26\pm13$ \kms.  This slight difference (at a low confidence level) suggests that weak BLA lines might be present in the \OVI\ systems and broadening their overall \HI\ profiles.  However, it is doubtful whether a difference between the two populations would be apparent without the \OVI\ detection ``sign-posts''.  There are several good, individual examples of this observational signature \citep[e.g.,][]{Tripp00,Tripp01,Stocke05,Stocke06}.

Observing multiple Lyman lines and determining \bHI\ rigorously through a COG tends to produce more accurate line width measurements, but even here, multiple components can broaden a line width.  Because the overestimate of the true $\btherm$ is smaller using a COG than for \Lya\ alone, we will adopt $\bCOG$ as our best estimator for absorber temperature.  Examining higher-order Lyman lines cannot be counted on to solve the multi-component problem since most BLA absorbers are too weak to allow \Lyb\ to be detected and measured in all components.  Nor will metal ions, which show intrinsically narrower lines, always help, since many BLAs are expected in regions of little or no metal enrichment.  Thus, we must adopt a statistical approach to BLA verification.
  
\begin{deluxetable*}{lccllll}
  \tabletypesize{\footnotesize}
  \tablecolumns{7} 
  \tablewidth{0pt} 
  \tablecaption{BLA Sample Sight Lines}
  \tablehead{\colhead{AGN}   &
           \colhead{R.A.}         &
	   \colhead{Decl.}        & 
	   \colhead{$z_{\rm AGN}$}&
           \colhead{$\Delta z_{\rm max}$} & 
           \colhead{$\Delta X_{\rm max}$} &
	   \colhead{Source \tablenotemark{a}}}
\startdata 
HE\,0226$-$4110  & 02:28:15.2  & $-$40:57:16 &  0.495  & 0.389 & 0.305 & \citet{Lehner06}  \\ 
HS\,0624$+$6907  & 06:30:02.5  & $+$69:05:04 &  0.370  & 0.354 & 0.282 & \citet{Aracil06a,Aracil06b} \\ 
PG\,1116$+$215   & 11:19:08.6  & $+$21:19:18 &  0.177  & 0.167 & 0.150 & \citet{Sembach04} \\ 
PG\,1259$+$593   & 13:01:12.9  & $+$59:02:07 &  0.478  & 0.388 & 0.303 & \citet{Richter04} \\ 
PKS\,0405$-$123  & 04:07:48.4  & $-$12:11:37 &  0.573  & 0.387 & 0.302 & \citet{Lehner07,Williger06}\\ 
H\,1821$+$643    & 18:21:57.3  & $+$64:20:36 &  0.297  & 0.283 & 0.236 & Sembach \etal, in prep.\\
PG\,0953$+$414   & 09:56:52.4  & $+$41:15:22 &  0.234  & 0.225 & 0.195 & Tripp \etal, in prep.\\
\enddata      
\tablenotetext{a}{Most recent detailed STIS/E140M analysis; all are also included in \citet{Lehner07,DS08}.}
\end{deluxetable*}
  
\subsection{Dataset and Methodology}

With the above points in mind, we set out to determine what fraction of the reported BLAs are legitimately tracing warm-hot gas at $T\ge10^5$~K.  To accomplish this, we have independently extracted, reduced, and scrutinized the STIS/E140M spectra (Table~1) used by \citet[][L07 hereafter]{Lehner07} to search for BLAs potentially arising in the WHIM.  Complete details of our data reduction method are given in DS08.  Briefly, STIS/E140M data were uniformly reduced using {\sc CalSTIS v2.19}.  Line-free continuum regions were defined interactively in 10\AA\ segments of the data and fitted with low-order Legendre polynomials.  Continuum fit uncertainty was taken as the standard deviation of points about the mean in the defined continuum region.  This uncertainty was added in quadrature with those from photon noise and line fit uncertainties.

The L07 work is the most complete and comprehensive look at BLAs currently available.  In order to determine the number and $b$-value distribution of BLA absorbers in the local universe, we have adopted the same definition as L07 and most other studies, requiring $T>10^5$~K corresponding to a hydrogen thermal $b>40$~\kms.  The relationship between COG-determined line width and line width measured from a single line shows a median ratio $\bLW/\bCOG=1.26^{+0.49}_{-0.25}$ in the large DS08 sample, and the correlation of that ratio with either \NHI\ or $\bLW$ is poor (Fig.~\ref{fig_DS08}).  Given this ratio, an absorber with $\bLW\ga50$~\kms\ (i.e., $1.26\times40$~\kms) would be more likely than not to have a $\bCOG\geq40$~\kms\ (see Figure~2).  Assuming that $\bCOG=\btherm$ (but, see discussion in Section 2.1).  We find that $T\geq10^5$~K for absorbers with line width $\bLW\ga50$~\kms, not the $\bLW\geq40$~\kms\ used by most previous works.  This difference is central to our analysis.  So {\em as a guideline}, we established the following categories and criteria for BLAs, using our analysis combined with that in the literature:

\noindent {\bf A. Probable BLA:} There is no definitive test for BLAs, but we classify as ``probable'' any system with well-defined $\bCOG\geq40$ \kms\ or $\bLW>60$ \kms\ confirmed by independent measurement of L07 and ourselves and with no obvious sign of multiple component structure (such as an asymmetric profile) in \Lya\ or higher Lyman lines or metal ions (if available).  Fifteen systems fall within this category.  Note that we do not {\it a priori} use the presence or absence of \OVI, \NV, or other WHIM sign posts as a BLA determinant, although several \OVI\ detections are present in this sample (see below).

\noindent {\bf B. Possible BLA:} Less likely than category A, the ``possible'' category includes systems with $40\leq\bLW\leq 60$~\kms.  This category also holds intermediate cases, where an absorber shows ambiguous component structure or a high degree of uncertainty as to line width and/or continuum fit, but a broad \HI\ system component cannot be reasonably ruled out.  Forty-eight \HI\ absorbers fall in this category.    

\begin{figure}
  \epsscale{1.2}\plotone{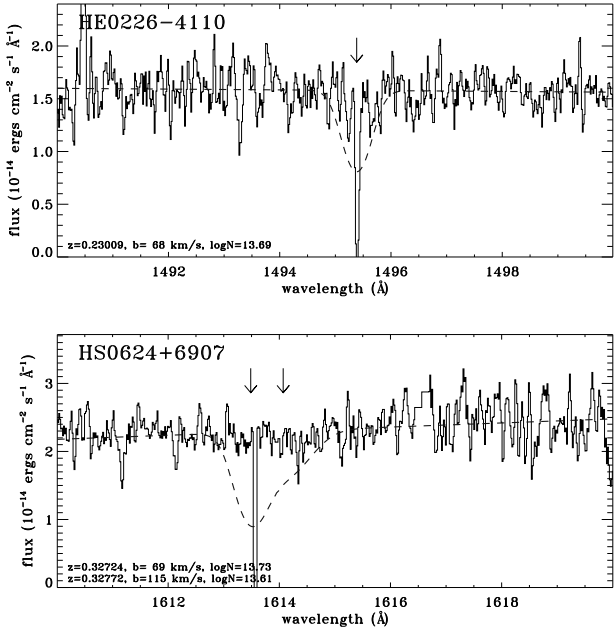} 
  \caption{Examples of three purported \Lya\ lines that do not appear
  in our reduction of the STIS/E140M data.  Reported redshifts,
  $b$-values, and column densities listed in L07 are printed in the
  lower left corner of each panel, and the resulting line profiles are
  overplotted.  Most of these ``missing'' lines are associated with
  very narrow bad-pixel regions where the flux drops to negative
  values in our version of the reduced data.}\label{fig_missing}
\end{figure}

\begin{figure}
  \epsscale{1.2}\plotone{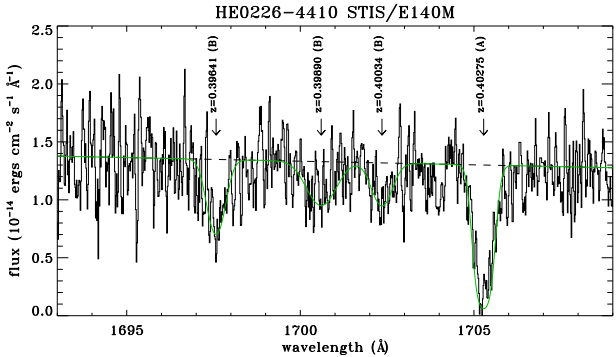}
  \caption{Potential ``echelle ripple'' behavior at the
  long-wavelength end of the STIS/E140M detector.  Four \Lya\
  absorbers are claimed \citep{Lehner06} but only the reddest
  ($z=0.40027$) is confirmed with a \Lyb\ counterpart.  While we view
  the other three absorbers with some skepticism, we find little or no
  similar ``echelle ripple'' behavior in other datasets and retain
  them as possible BLAs in our accounting.}  \label{fig_echripple}
\end{figure}

\noindent {\bf C. Non-BLA:} Nearly half (56/119) of the \HI\ lines with $b\geq40$~\kms\ published by L07 or DS08 are most likely {\em not} BLAs by our assessment for a variety of reasons:  measured $\bCOG<40$ \kms\ or $\bLW<40$ \kms; probable alternate line identification; obvious narrow component structure; or simply not detected as absorption features in our reduction of the data (i.e., we do not confirm the extraction and/or analysis done by or referenced in L07, see Figure~\ref{fig_missing}).  The last group, encompassing nine purported absorbers, occurs primarily in the spectrum of HS\,0624$+$6907 \citep{Aracil06a,Aracil06b}.  In the course of our independent analysis, we identified a few BLA candidates that were not listed in L07 or other independent literature sources.  In most cases, these lines turn out to be something other than \Lya\ and thus fall into category C, as they are not confirmed by two independent sources.  

Because the 7 individual sightlines were analyzed originally by different authors, somewhat differing analysis procedures were used and then adopted by L07.  Each is briefly discussed below, with notes that bear on the BLA identification process.

\paragraph{HE\,0226$-$4110}
Originally analyzed by \citet{Lehner06}, the spectrum is shown in that paper.  No COG measurements are reported; instead some $b$-values are obtained by simultaneous fitting of several Lyman-series lines (see discussion below).  Longward of 1690 \AA, the continuum in this spectrum appears to ``ripple'' in both our extracted spectrum (Figure~\ref{fig_echripple})  that of Lehner \etal.  This gives rise to four reported BLA candidates (one with $\bLW\sim150$~\kms) in only $\Delta z=0.007$.  The periodic nature of these features are suspicious, especially given that these features occur the highest spectral orders which cut across many detecor rows before extraction.  However, no similar features appear in the highest spectral order ($1690<\lambda<1710$\AA) for AGN where $z_{\rm Ly\alpha}>z_{\rm AGN}$.

\paragraph{HS\,0624$+$6907}
Originally analyzed by \citet{Aracil06a,Aracil06b}, no spectra were published.  These authors do not mention obtaining COG $b$-values except for a \Lya\ absorption complex at $z\sim0.06$.  Therefore, we assume that all $b$-values quoted by L07 for this spectrum are either \Lya\ line width values or simultaneous line width fits to two or more Lyman lines as described for HE\,0226$-$4110 above.  Several of the BLAs reported by L07 in this sight line are at or near the locations of detector artifacts;  we see no believable absorption at these locations based on our own reduction of the data.  The reported lines may be a result of smoothing the data over these bad pixels.  Longward of 1580  \AA\ in this spectrum our extracted data does {\it not} match even the presence of some of the absorption lines listed by L07 (Figure~\ref{fig_missing}).  These unconfirmed lines are listed in Table~2 as non-BLAs for completeness, but are not otherwise analyzed.  

\paragraph{PG\,1116$+$215}
Originally analyzed by \citet{Sembach04}, the spectrum is shown in that paper. \citet{Sembach04} use COG $b$-values where available, although none apply to the BLAs in this sight line.

\paragraph{PG\,1259$+$593}
This sight line was originally analyzed by \citet{Richter04} and the spectrum is shown in that paper.  Richter \etal\ list several COG $b$-values for BLAs and we add several more.

\paragraph{PKS\,0405$-$123}
Originally analyzed by \citet[W06]{Williger06} and reanalyzed by L07.  Neither W06 nor L07 report COG measurements for this sightline.  The spectrum is shown in Williger \etal\  Despite mentioning that the reanalysis was required due to the large number of suspect BLA candidates, L07 $b$-values from simultaneous line fits are similar to the \Lya-only line-widths reported by \citet{Williger06}.  We report several COG $b$-values for this sight line and, similar to many other absorption systems we generally find $\bCOG\leq\bLW$ (see below). 

\paragraph{H\,1821$+$643}
For spectral analysis, L07 refer to Sembach \etal\ In prep. which does not appear to have been published yet.  There is no mention of COG measurements in the L07 description, so we have assumed that the reported values are $\bLW$.

\paragraph{PG\,0953$+$415}
L07 report these BLAs based upon an unpublished Tripp \etal\ analysis.  We have assumed that all quoted $b$-values are based on \Lya\ line width measurements since our own $\bLW$ values are similar.

One of the procedural differences between L07 and the current work is that where more than one Lyman line is detected in absorption in these spectra, we have used a $b$-value determined by the COG method whereas L07 uses a simultaneous line fit of all available lines.  In general, the $\bCOG$ values are significantly smaller than those obtained by the simultaneous line fitting method.  To evaluate this difference, we have used the 10 absorbers in the PKS\,0405$-$123 sight line analyzed independently by ourselves, L07, and W06.  For all ten systems the W06 ($\bLW$) measurements agree to within the errors with the L07 (simultaneous line-fits) $b$-values.  Half of these measurements also show excellent agreement with $\bCOG$.  However, the other five systems show $\bCOG$ values significantly less than values obtained by the simultaneous line fit method (for details see Table~2) and these disagreements are for the largest $b$-values reported by L07 in this sightline.  Three systems ($z=0.09659$, $0.16121$, and $0.32500$) have COG $b$-values with tight error bars half or less of the L07 reported values;  two other systems ($z=0.17876$ and $0.25861$) have similar differences between reported values but, with large $\bCOG$ error bars and so are not inconsistent with the large line widths reported by L07.  Based upon these 10 systems only, we find significant differences between COG $b$-values and simultaneous line-fit $b$-values in roughly half the cases, all in the sense that the COG $b$-values are significantly less (35-50\% the amount).  When higher S/N FUV spectra from COS become available, tests using these two techniques should be made to determine which method is the most reliable as a function of line width and S/N. 

\section{Results}

\begin{figure*}
  \epsscale{1.15}\plotone{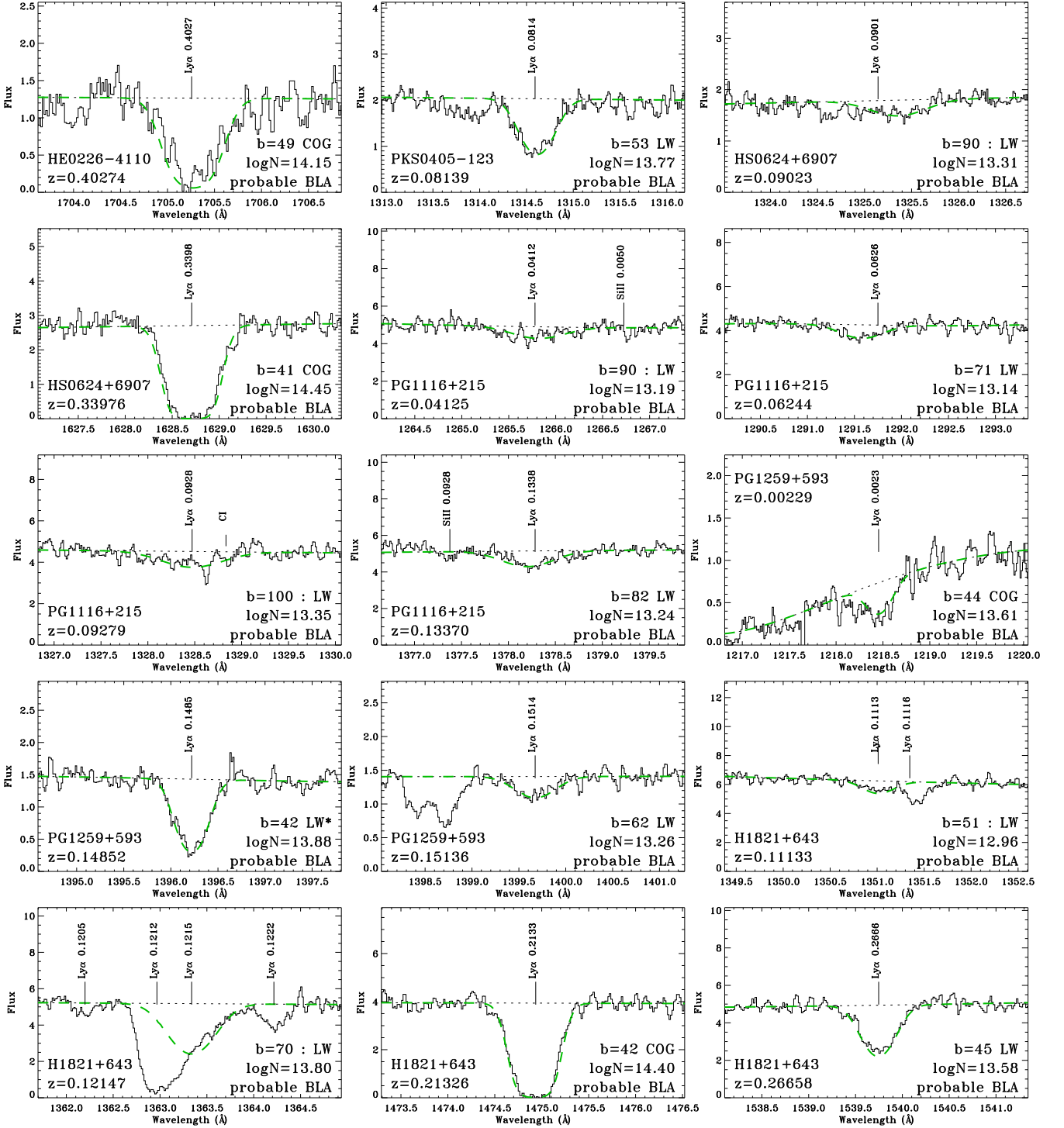} 
  \caption{Spectra of all 15 Probable BLA systems illustrate the
  diversity of line profiles.  The sight line and absorber redshift
  are listed on the left side of each panel while the consensus $b$
  (\kms) and $\log\,N$ values from Table~2 are listed on the right.
  Dotted line shows continuum fit to the data, and dashed curve shows
  consensus absorption profile.  Adjacent line detections are
  indicated with vertical ticks and labeled with ion and redshift.
  Lines identified without redshift denote Galactic interstellar
  absorption.  Flux is in units of
  $10^{-14}\rm~erg~cm^{-2}~s^{-1}~\AA^{-1}$.  See Table~2 and Appendix
  for more details of individual systems.}  \label{fig_examples1}
\end{figure*}

\begin{figure*}
  \epsscale{1.15}\plotone{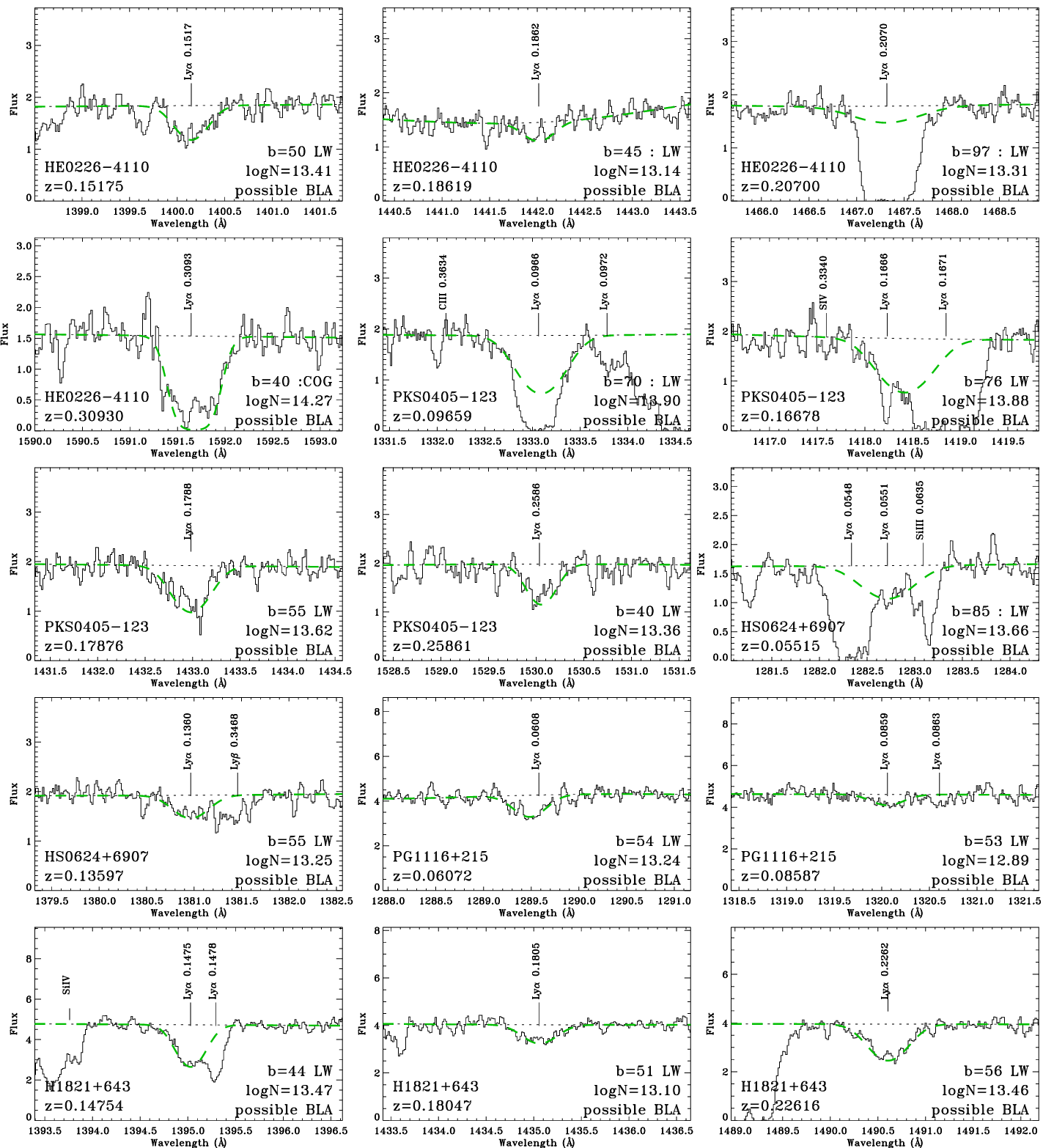} 
  \caption{Same as Fig.~\ref{fig_examples1} but for 15 of the 48
  Possible BLA systems.  Additional Possible BLAs are shown in
  Figure~\ref{fig_cos}.}
  \label{fig_examples2}
\end{figure*}

All BLA candidates drawn from the catalogs of L07 and DS08 were independently analyzed by both Danforth and Stocke based on the criteria above.  In Table~2, we list the Probable (A), Possible (B), and non-BLA (C) absorbers based upon our analyses.  Sight line name and absorber redshift are in the first two columns.  Independent $b$-values and column densities are listed from L07 (or similar source, columns 3-5) and DS08/this work (columns 6-8).  COG measurements of \bHI, \NHI\ are given where possible, otherwise the unweighted mean of the two independent \Lya\ measurements is taken as consensus $b$, \NHI\ values along with the method used (`LW' for \Lya\ line width; `COG' for curve of growth) in columns 9 and 10.  Brief notes are given for each absorber, with additional details presented in an Appendix for many absorbers in column 11.  In total, we find 15 absorbers in group A (Probable BLAs) and 48 in group B (Possible BLAs) and 56 non-BLAs out of 119 candidate systems with $b>40$~\kms\ as reported by L07, DS08, or similar literature source.  We show the Probable BLAs in Figure~\ref{fig_examples1} and a selection of Possible BLAs in Figure~\ref{fig_examples2}.

\LongTables
\begin{deluxetable*}{llccccccccl}
\tabletypesize{\footnotesize}
\tablecolumns{11} 
\tablewidth{0pt} 
\tablecaption{Probable, Possible, and non-BLA Absorbers}
\tablehead{\colhead{Sight Line}    &
   \colhead{$z_{\rm abs}$}         &
   \colhead{$b_1$}         &
   \colhead{log\,$N_1$}    &
   \colhead{Src\tnma}        &
   \colhead{$b_2$}         &
   \colhead{log\,$N_2$}    &
   \colhead{Src\tnma}        &
   \colhead{$b_0$}         &
   \colhead{log\,$N_0$}    &
   \colhead{Absorber Notes}        \\ 
\colhead{}                         &
   \colhead{}         &
   \colhead{(\kms)}   &
   \colhead{(cm$^{-2}$)}    &
   \colhead{1}        &
   \colhead{(\kms)}   &
   \colhead{(cm$^{-2}$)}    &
   \colhead{2}        &
   \colhead{(\kms)}   &
   \colhead{(cm$^{-2}$)}    &
   \colhead{}                      }
\startdata 
 \cutinhead{Probable BLA Detections}
HE\,0226$-$4110& 0.40274 &$46\pm4$ &$14.13\pm0.04$&L07 &$49$ COG&$14.16\pm0.11$&DS08 &49 COG   &14.15&\Lya,$\beta$ detections \\ 
PKS\,0405$-$123& 0.08139 &$54\pm4$ &$13.79\pm0.02$&L07 &$53\pm2$&$13.76\pm0.02$&DS08 &53 LW    &13.77&see appendix \\ 
HS\,0624$+$6907& 0.09023 &$76\pm14$ &$13.29\pm0.08$&L07 &$100:$&$13.34\pm0.04$&this &90: LW   &13.31&very weak and broad \\ 
HS\,0624$+$6907& 0.33976 &$42\pm1$ &$14.45\pm0.03$&L07 &$41$ COG&$14.46^{+0.10}_{-0.03}$&DS08 &41 COG   &14.45&OVI, see appendix \\ 
PG\,1116$+$215 & 0.04125 &$105\pm18$&$13.25^{+0.11}_{-0.09}$&L07 &$72\pm6 $&$13.13^{+0.06}_{-0.05}$&DS08 &90: LW   &13.19&very weak \\ 
PG\,1116$+$215 & 0.06244 &$77\pm9$ &$13.18^{+0.07}_{-0.06}$&L07 &$65\pm7 $&$13.10\pm0.06$&this &71  LW   &13.14&very weak \\ 
PG\,1116$+$215 & 0.09279 &$133\pm17$&$13.39^{+0.09}_{-0.08}$&L07 &$83:        $&$13.31^{+0.04}_{-0.03}$&this &100: LW   &13.35&see appendix \\ 
PG\,1116$+$215 & 0.13370 &$84\pm10$ &$13.27^{+0.08}_{-0.07}$&L07 &$79\pm6 $&$13.20\pm0.04$&DS08 &82  LW   &13.24&OVI, see appendix \\ 
PG\,1259$+$593 & 0.00229 &$44^{+9}_{-4}$ COG&$13.61\pm0.06$&R04 &$61\pm5 $&$13.83\pm0.24$&this &44 COG   &13.61&see appendix \\ 
PG\,1259$+$593 & 0.14852 &$42\pm2$&$13.91\pm0.06$&L07 &$57:$ COG&$13.85\pm0.01$&this &42 LW*   &13.88&see appendix \\ 
PG\,1259$+$593 & 0.15136 &$65\pm6$&$13.32\pm0.09$&L07 &$59\pm5 $&$13.21\pm0.05$&DS08 &62 LW    &13.26& \\ 
H\,1821$+$643  & 0.11133 &$88\pm14$&$12.95^{+0.10}_{-0.13}$&L07 &$51:$        &$12.97\pm0.03$&this &51: LW   &12.96&see appendix \\ 
H\,1821$+$643  & 0.12147 &$40^{+44}_{-21}$&$14.04\pm0.36$&L07 &$76\pm5 $&$13.79\pm0.06$&this &70: LW   &13.80&OVI, see appendix \\ 
H\,1821$+$643  & 0.21326 &$43\pm2$&$14.41^{+0.04}_{-0.04}$&L07 &$42^{+5}_{-4}$ COG&$14.40^{+0.08}_{-0.09}$&DS08 &42 COG   &14.40&OVI, see appendix \\ 
H\,1821$+$643  & 0.26658 &$45\pm2$&$13.64^{+0.03}_{-0.03}$&L07 &$45\pm1 $&$13.52\pm0.01$&DS08 &45 LW    &13.58&OVI, see appendix \\ 
\cutinhead{Possible BLA Detections}
HE\,0226$-$4110& 0.06083 &$45\pm1$&$14.65\pm0.02$&L07 &$46\pm15 $&$14.86\pm0.15$&this &45 LW    &14.75&see appendix \\ 
HE\,0226$-$4110& 0.09220 &$40\pm18$&$12.94\pm0.11$&L07 &$47\pm9 $&$13.11^{+0.08}_{-0.05}$&this &44 LW    &13.03&weak, profile uncertain \\ 
HE\,0226$-$4110& 0.15175 &$49\pm7$&$13.42\pm0.05$&L07 &$51\pm5 $&$13.39\pm0.03$&this &50 LW    &13.41&asymmetric \\ 
HE\,0226$-$4110& 0.16339 &$46\pm2$&$14.36\pm0.04$&L07 &$39\pm7 $&$14.34\pm0.03$&this &42 LW    &14.35&see appendix \\ 
HE\,0226$-$4110& 0.18619 &$54\pm16$&$13.26\pm0.08$&L07 &$37\pm8 $&$13.02\pm0.07$&this &45: LW   &13.14&possibly two components \\ 
HE\,0226$-$4110& 0.20700 &$97:  $&$13.31\pm0.39$&S05 &blend&$<15.20$&this &97: LW   &13.31&OVI, NeVIII, see appendix \\ 
HE\,0226$-$4110& 0.30930 &$44\pm2$&$14.26\pm0.03$&L07 &$38^{+21}_{-9}$ COG&$14.28^{+0.20}_{-0.17}$&this &40:COG   &14.27&see appendix \\ 
HE\,0226$-$4110& 0.39641 &$63\pm23$&$13.59\pm0.10$&L07 &$57\pm10 $&$13.54\pm0.10$&this &60 LW    &13.57&see appendix \\ 
HE\,0226$-$4110& 0.39890 &$152: $&$13.50\pm0.16$&L07 &$88\pm20 $&$13.46\pm0.10$&this &100: LW   &13.50&see appendix \\ 
HE\,0226$-$4110& 0.40034 &$61\pm26$&$13.39\pm0.11$&L07 &$74\pm16 $&$13.29\pm0.11$&this &70 LW    &13.34&see appendix \\ 
PKS\,0405$-$123& 0.03196 &$54\pm16$&$13.33\pm0.08$&L07 &$57\pm8 $&$13.33\pm0.06$&DS08 &56 LW    &13.33&see appendix \\ 
PKS\,0405$-$123& 0.07523 &$48\pm20$&$13.05\pm0.11$&L07 &$37\pm12 $&$12.81\pm0.13$&this &43 LW    &12.93&see appendix\\
PKS\,0405$-$123& 0.09659 &$70\pm20$&$13.90\pm0.18$&L07 &$36\pm4$ COG&$14.64^{+0.12}_{-0.11}$&DS08 &70: LW   &13.90&OVI, see appendix \\ 
PKS\,0405$-$123& 0.13102 &$52\pm8$&$13.46\pm0.05$&L07 &$53\pm6 $&$13.35\pm0.04$&this &57 LW    &13.41& see Appendix \\ 
PKS\,0405$-$123& 0.13377 &$43\pm8$&$13.34\pm0.06$&L07 &$45\pm7 $&$13.22\pm0.07$&this &44 LW    &13.28& see Appendix \\ 
PKS\,0405$-$123& 0.16678 &$75\pm7$&$13.91\pm0.04$&L07 &$77\pm7 $&$13.84\pm0.05$&this &76 LW    &13.88&poss. OVI, see appendix \\ 
PKS\,0405$-$123& 0.17876 &$55\pm7$&$13.61\pm0.04$&L07 &$18^{+44}_{-7}$ COG&$13.92^{+0.29}_{-0.31}$&DS08 &55 LW    &13.62&asymetric, see appendix \\ 
PKS\,0405$-$123& 0.18269 &$48\pm2$ COG&$15.07\pm0.09$&W06 &$49^{+10}_{-6}$ COG&$14.90^{+0.15}_{-0.20}$&DS08 &49 COG   &14.86&OVI, see appendix \\ 
PKS\,0405$-$123& 0.19086 &$44\pm16$&$13.17\pm0.09$&L07 &$38\pm9 $&$13.01^{+0.10}_{-0.06}$&this &41 LW    &13.09& see Appendix \\
PKS\,0405$-$123& 0.24513 &$54\pm24$&$13.23\pm0.11$&L07 &$30\pm5 $&$13.01\pm0.09$&this &42: LW   &13.12&poor agreement, see appendix \\ 
PKS\,0405$-$123& 0.25861 &$40\pm9$&$13.37\pm0.07$&L07 &$39\pm4 $&$13.36^{+0.15}_{-0.10}$&this &40 LW    &13.36&$\bCOG=21^{+25}_{-10}$; W06:$\bLW=47\pm7$\\ 
PKS\,0405$-$123& 0.29523 &$47\pm13$&$13.33\pm0.08$&L07 &$39\pm8 $&$13.07\pm0.08$&this &43 LW    &13.20& \\ 
PKS\,0405$-$123& 0.29904 &$49\pm23$&$13.26\pm0.12$&L07 &$54\pm10 $&$13.14\pm0.15$&this &52: LW   &13.20&noisy, components? \\ 
PKS\,0405$-$123& 0.35092 &$38\pm2$&$14.25\pm0.03$&L07 &$56\pm15$ COG&$14.04\pm0.08$&DS08 &40: LW   &14.25&see appendix \\ 
PKS\,0405$-$123& 0.40886 &$40\pm2$&$14.38\pm0.03$&L07 &$40^{+7}_{-4}$ COG&$14.32\pm0.05$&DS08 &40:COG   &14.35&see appendix \\ 
HS\,0624$+$6907& 0.05437 &$60\pm19$&$13.09\pm0.11$&L07 &$45\pm5 $&$12.88^{+0.13}_{-0.11}$&this &52: LW   &12.99&see appendix \\ 
HS\,0624$+$6907& 0.05515 &$84\pm31$&$13.68\pm0.17$&L07 &$87\pm13 $&$13.69\pm0.06$&this &85: LW   &13.66&see appendix \\ 
HS\,0624$+$6907& 0.13597 &$57\pm11$&$13.33\pm0.10$&L07 &$52\pm12 $&$13.16\pm0.10$&this &55  LW   &13.25&OVI, see appendix \\ 
HS\,0624$+$6907& 0.21323 &$45\pm6$&$13.22\pm0.05$&L07 &$40\pm7 $&$13.04\pm0.08$&DS08 &43  LW   &13.13&poss. SiIII, CIII \\ 
HS\,0624$+$6907& 0.26856 &$51\pm7$&$13.03\pm0.05$&L07 &$46\pm8 $&$13.00\pm0.07$&this &49  LW   &13.02&weak \\ 
HS\,0624$+$6907& 0.29661 &$52\pm3$&$13.54\pm0.02$&L07 &$45         $&$13.31\pm0.04$&DS08 &48  LW   &13.43& \\ 
HS\,0624$+$6907& 0.30994 &$66\pm12$&$13.61\pm0.10$&L07 &$38\pm9 $&$12.90\pm0.11$&DS08 &52: LW   &13.25&poss. OVI, see appendix \\ 
HS\,0624$+$6907& 0.31790 &$34\pm4$&$13.37\pm0.04$&L07 &$51\pm7 $&$13.38^{+0.07}_{-0.05}$&DS08 &43 LW    &13.38&OVI, see appendix \\ 
PG\,0953$+$415 & 0.05879 &$63^{+19}_{-14}$&$13.41\pm0.16$&L07 &not measured&not measured&\nodata  &63: LW   &13.37&see appendix \\ 
PG\,0953$+$415 & 0.17985 &$48^{+11}_{-9}$&$13.27\pm0.07$&L07 &$48\pm2 $&$13.20^{+0.07}_{-0.05}$&DS08 &48 LW    &13.24&see appendix \\ 
PG\,0953$+$415 & 0.19126 &$48^{+95}_{-32}$&$13.08\pm0.55$&L07 &$66:        $&$13.50\pm0.04$&this &40: LW   &13.30&see appendix \\ 
PG\,0953$+$415 & 0.20104 &$71^{+41}_{-26}$&$13.16\pm0.16$&L07 &$44:        $&$12.63\pm0.19$&this &57: LW   &12.90&weak, uncertain continuum \\ 
PG\,1116$+$215 & 0.01638 &$49\pm5$&$13.39\pm0.06$&L07 &$54\pm4 $&$13.44\pm0.03$&DS08 &51 LW    &13.41&asymmetric \\ 
PG\,1116$+$215 & 0.06072 &$55\pm6$&$13.28^{+0.06}_{-0.05}$&L07 &$53         $&$13.21\pm0.04$&DS08 &54 LW    &13.24& \\ 
PG\,1116$+$215 & 0.08587 &$52\pm14$&$12.90^{+0.19}_{-0.13}$&L07 &$55\pm11 $&$12.88\pm0.08$&this &53 LW    &12.89&see appendix \\ 
PG\,1259$+$593 & 0.19573 &$43\pm7$ &$13.07\pm0.14$&R04 &$46\pm19 $&$13.05\pm0.15$&DS08 &44 LW    &13.06&see appendix \\ 
PG\,1259$+$593 & 0.41786 &$51\pm4$ &$13.25\pm0.08$&L07 &$52\pm24 $&$13.23\pm0.18$&DS08 &51: LW   &13.24&asymmetric \\ 
H\,1821$+$643  & 0.02642 &$49\pm6$ &$13.26^{+0.07}_{-0.08}$&L07 &$44\pm3 $&$13.19\pm0.03$&DS08 &47 LW    &13.23&asymmetric \\ 
H\,1821$+$643  & 0.14754 &$45\pm3$ &$13.51^{+0.03}_{-0.03}$&L07 &$42\pm2 $&$13.43\pm0.02$&this &44 LW    &13.47&see appendix \\ 
H\,1821$+$643  & 0.16352 &$52\pm6$ &$13.17^{+0.06}_{-0.07}$&L07 &$56\pm4 $&$13.11\pm0.04$&DS08 &54 LW    &13.14&weak, slightly asymmetric \\ 
H\,1821$+$643  & 0.18047 &$51\pm7$ &$13.14^{+0.07}_{-0.08}$&L07 &$52\pm8 $&$13.05\pm0.07$&DS08 &51 LW    &13.10& \\ 
H\,1821$+$643  & 0.22616 &$55\pm4$ &$13.51^{+0.04}_{-0.04}$&L07 &$56\pm2 $&$13.41\pm0.07$&this &56 LW    &13.46&offset OVI, see appendix \\ 
H\,1821$+$643  & 0.25814 &$60\pm9$ &$13.38^{+0.10}_{-0.13}$&L07 &$56:        $&$13.26\pm0.04$&this &58: LW   &13.32&see appendix \\ 
 \cutinhead{Not BLAs}
HE\,0226$-$4110& 0.01216 &not detected&not detected&\nodata &$41\pm7 $&$13.06^{+0.08}_{-0.07}$&DS08 &\nodata&\nodata&poor S/N, not confirmed \\ 
HE\,0226$-$4110& 0.02679 &$42\pm11$&$13.22\pm0.08$&L07 &$30\pm4 $&$13.17^{+0.12}_{-0.05}$&DS08 &36 LW    &13.19&Lehner et al. 2006:$\bLW=28$ \\ 
HE\,0226$-$4110& 0.23009 &$68\pm8$ &$13.69\pm0.04$&L07 &not detected&not detected&\nodata  &\nodata&\nodata&no absorption seen \\ 
HE\,0226$-$4110& 0.22102 &$34\pm18$&$12.99\pm0.12$&L07 &$44\pm9 $&$13.06^{+0.07}_{-0.08}$&this &39 LW    &13.02&weak \\ 
HE\,0226$-$4110& 0.38420 &$62\pm7$ &$13.91\pm0.04$&L07 &$31:$ COG&$13.08^{+0.11}_{-0.10}$&this &31: COG  &13.49&components, see appendix \\ 
PKS\,0405$-$123& 0.05896 &$77\pm27$&$13.34\pm0.10$&L07 &$20, 20     $&$12.97\pm0.08$&this &20 LW    &12.97& see appendix \\
PKS\,0405$-$123& 0.07218 &$45\pm14$&$13.09\pm0.08$&L07 &$45:        $&$13.01\pm0.06$&this &45 LW    &13.05&see appendix \\ 
PKS\,0405$-$123& 0.10298 &$87\pm19$&$13.40\pm0.07$&L07 &$54:        $&$13.35^{+0.04}_{-0.03}$&this &70: LW   &13.38&see appendix \\ 
PKS\,0405$-$123& 0.10419 &not detected&not detected&\nodata &$48\pm5 $&$12.98\pm0.08$&DS08 &\nodata&\nodata& see appendix \\
PKS\,0405$-$123& 0.13646 &$54\pm11$&$13.34\pm0.06$&L07 &$11, 19     $&$12.96\pm0.07$&this &19: LW   &12.96& see appendix \\
PKS\,0405$-$123& 0.13924 &$35\pm11$&$13.02\pm0.11$&W06 &$53\pm2 $&$13.13^{+0.06}_{-0.05}$&DS08 &44 LW    &13.08& see appendix \\
PKS\,0405$-$123& 0.15304 &$46\pm3$ &$13.80\pm0.03$&L07 &$50:        $&$13.76\pm0.03$&this &48: LW   &13.78&see appendix \\ 
PKS\,0405$-$123& 0.16121 &$54\pm8$ &$13.71\pm0.04$&L07 &$27\pm14$ COG&$13.62^{+0.31}_{-0.15}$&DS08 &27 COG   &13.66& see appendix \\ 
PKS\,0405$-$123& 0.16714 &$30\pm1$ &$16.27\pm0.13$&L07 &$40^{+12}_{-7}$ COG&$15.47^{+0.40}_{-0.48}$&DS08 &\nodata&\nodata&OVI, components, see appendix \\ 
PKS\,0405$-$123& 0.24057 &$57\pm19$&$13.27\pm0.09$&L07 &$58\pm6 $&$13.27^{+0.08}_{-0.07}$&this &58 LW    &13.27&marginal feature, artifact \\ 
PKS\,0405$-$123& 0.28838 &$52\pm19$&$13.32\pm0.10$&L07 &$30:        $&$13.18\pm0.10$&this &30: LW   &13.10&components; W06:$\bLW=27\pm7$ \\ 
PKS\,0405$-$123& 0.32500 &$66\pm13$&$13.55\pm0.06$&L07 &$60:        $&$13.45^{+0.04}_{-0.08}$&this &63: LW   &13.50&see appendix \\ 
PKS\,0405$-$123& 0.34234 &$42\pm13$&$13.39\pm0.08$&L07 &$35\pm4 $&$13.27^{+0.07}_{-0.05}$&this &38  LW   &13.33&asymetric \\ 
PKS\,0405$-$123& 0.36150 &$44\pm10$&$13.71\pm0.10$&L07 &$25\pm3$ COG&$15.18\pm0.13$&this &30: COG  &15.00&see appendix \\ 
HS\,0624$+$6907& 0.04116 &$41\pm3$ &$13.33\pm0.03$&L07 &not measured&not measured&\nodata  &\nodata&\nodata&SIII$z=0.0635$, see appendix \\ 
HS\,0624$+$6907& 0.05483 &$35:       $&$14.50:$&L07 &$45^{+28}_{-9}$ COG&$14.28^{+0.11}_{-0.15}$&DS08 &35: LW   &14.28&see appendix \\ 
HS\,0624$+$6907& 0.06346 &$48\pm8$ &$14.46\pm0.30$&L07 &$33^{+2}_{-1}$ COG&$15.25^{+0.10}_{-0.06}$&DS08 &33 COG   &15.25&OVI, see appendix \\ 
HS\,0624$+$6907& 0.19979 &$17\pm2$ &$13.24\pm0.05$&L07 &$55\pm10 $&$13.58\pm0.03$&DS08 &17 LW    &13.41&poss. OVI; components in \Lyb \\ 
HS\,0624$+$6907& 0.21990 &$60\pm9$ &$13.39\pm0.05$&L07 &not detected&not detected&\nodata  &\nodata&\nodata&not detected \\ 
HS\,0624$+$6907& 0.23231 &$44\pm8$ &$13.33\pm0.08$&L07 &not detected&not detected&\nodata  &\nodata&\nodata&not detected \\ 
HS\,0624$+$6907& 0.28017 &$43\pm2$ &$14.32\pm0.02$&L07 &$35^{+6}_{-5}$ COG&$14.42^{+0.09}_{-0.07}$&DS08 &35 COG   &14.37& \\ 
HS\,0624$+$6907& 0.29531 &$42\pm2$ &$13.80\pm0.02$&L07 &$34         $&$13.62\pm0.03$&DS08 &38 LW    &13.71& \\ 
HS\,0624$+$6907& 0.31045 &$62\pm40$&$13.43\pm0.33$&L07 &not detected&not detected&\nodata  &\nodata&\nodata&see appendix \\ 
HS\,0624$+$6907& 0.31088 &$51\pm28$&$13.13\pm0.43$&L07 &not detected&not detected&\nodata  &\nodata&\nodata&see appendix \\ 
HS\,0624$+$6907& 0.31280 &$54\pm9$ &$13.65\pm0.10$&L07 &not detected&not detected&\nodata  &\nodata&\nodata&not detected \\ 
HS\,0624$+$6907& 0.31326 &$55\pm11$&$13.62\pm0.10$&L07 &not detected&not detected&\nodata  &\nodata&\nodata&not detected \\ 
HS\,0624$+$6907& 0.32089 &$31\pm1$ &$13.97\pm0.02$&L07 &$44^{+13}_{-14}$ COG&$13.80^{+0.06}_{-0.11}$&DS08 &31 LW    &13.89&see appendix \\ 
HS\,0624$+$6907& 0.32724 &$69\pm16$&$13.73\pm0.32$&L07 &not detected&not detected&\nodata  &\nodata&\nodata&not detected \\ 
HS\,0624$+$6907& 0.32772 &$115\pm62$&$13.61\pm0.43$&L07 &not detected&not detected&\nodata  &\nodata&\nodata&not detected \\ 
PG\,0953$+$415 & 0.02336 &$56^{+14}_{-11}$&$13.21\pm0.08$&L07 &$20:        $&$12.68^{+0.19}_{-0.12}$&this &38: LW   &12.95&discrepant measurements \\ 
PG\,0953$+$415 & 0.04382 &not detected&not detected&\nodata &$47\pm8 $&$13.39^{+0.03}_{-0.04}$&DS08 &47 LW    &13.06&OVI$z=0.22974$; see appendix \\ 
PG\,0953$+$415 & 0.12784 &$44:       $&$12.83\pm0.35$&L07 &not detected&not detected&this &\nodata&\nodata&not detected, see appendix \\ 
PG\,0953$+$415 & 0.19361 &$40\pm2$ &$13.94\pm0.02$&L07 &$37\pm2 $&$13.82\pm0.03$&this &39 COG   &14.15&see appendix \\ 
PG\,0953$+$415 & 0.20006 &$66^{+18}_{-14}$&$13.24\pm0.09$&L07 &$32\pm7 $&$12.91\pm0.11$&DS08 &49: LW   &13.07&continuum uncertainties \\ 
PG\,1116$+$215 & 0.02841 &$31\pm1$ &$13.80\pm0.02$&L07 &$43:        $&$13.71^{+0.21}_{-0.09}$&DS08 &35 LW    &13.78& \\ 
PG\,1116$+$215 & 0.05904 &$21,30     $&???&S04 &$45:        $&$13.53\pm0.02$&DS08 &21,30 LW &13.53&components \\ 
PG\,1259$+$593 & 0.04606 &$48\pm12$&$15.58\pm0.21$&L07 &$30^{+10}_{-8}$ COG&$15.51^{+0.28}_{-0.25}$&DS08 &32 COG   &15.55&OVI, components, see appendix \\ 
PG\,1259$+$593 & 0.06931 &not detected&not detected&\nodata &$54\pm4 $&$13.33\pm0.04$&DS08 &54 LW    &13.33&not detected by R04 \\ 
PG\,1259$+$593 & 0.08041 &$42\pm4$ &$12.97\pm0.10$&L07 &$26\pm2 $&$12.86$&this &34: LW   &12.92&not detected by R04 \\ 
PG\,1259$+$593 & 0.17891 &$99\pm9$ &$13.29\pm0.10$&L07 &not measured&not measured&\nodata  &\nodata&\nodata&not detected \\ 
PG\,1259$+$593 & 0.21136 &not measured&not measured&\nodata &$47\pm3 $&$13.39\pm0.04$&DS08 &47 LW    &13.39&see appendix \\ 
PG\,1259$+$593 & 0.22861 &$40\pm3$ &$13.47\pm0.05$&L07 &$34\pm2 $&$13.42\pm0.03$&DS08 &37 LW    &13.44& \\ 
PG\,1259$+$593 & 0.24126 &$89\pm7$ &$13.41\pm0.09$&L07 &$23\pm4 $&$12.86\pm0.08$&this &55 LW    &13.14&ambiguous components \\ 
PG\,1259$+$593 & 0.25971 &$41\pm5$ &$13.84\pm0.12$&L07 &$29^{+7}_{-5}$ COG&$13.98\pm0.07$&DS08 &29 COG   &13.91&OVI, obvious components \\ 
PG\,1259$+$593 & 0.30434 &$65\pm1$ &$13.76\pm0.14$&L07 &$20^{+8}_{-3}$ COG&$13.81^{+0.07}_{-0.09}$&DS08 &20 COG   &13.81&obvious components \\ 
PG\,1259$+$593 & 0.31978 &$74\pm8$ &$13.98\pm0.06$&L07 &$31\pm3$ COG &$14.07^{+0.01}_{-0.06}$&DS08 &31 COG   &14.07&OVI; listed as blend in R04 \\ 
PG\,1259$+$593 & 0.32478 &$46\pm10$&$13.24\pm0.15$&L07 &$20 $&$12.93^{+0.14}_{-0.11}$&DS08 &33: LW   &13.10&OVI; continuum uncertainty? \\ 
PG\,1259$+$593 & 0.43569 &$44\pm4$ &$14.22\pm0.10$&R04 &\nodata&\nodata&\nodata  &\nodata&\nodata&see appendix \\ 
H\,1821$+$643  & 0.12221 &$42\pm4$ &$13.19\pm0.06$&L07 &$34\pm5 $&$13.13^{+0.03}_{-0.04}$&DS08 &38 LW    &13.16& \\ 
H\,1821$+$643  & 0.19176 &not detected&not detected&\nodata &$47\pm7 $&$12.72\pm0.07$&DS08 &\nodata&\nodata&marginal; not confirmed by L07 \\ 
H\,1821$+$643  & 0.22480 &not detected&not detected&\nodata &$40^{+6}_{-8}$ COG&$15.41^{+0.25}_{-0.08}$&DS08 &40 COG   &15.41&OVI; components, see appendix \\ 
  \enddata
  \tablenotetext{a}{Measurement sources: L07 (Lehner et al. 2007 and sources therein); DS08 (Danforth \& Shull 2008); W06 (Williger et al. 2006); S05 (Savage et al. 2005); S04 (Sembach et al. 2004); R04 (Richter et al. 2004); this (this work).}
  \label{tab:blatable}
\end{deluxetable*}

\subsection{Broad \Lya\ Absorber Frequency}

In measuring metal-ion absorption lines, DS08 employ a $4\sigma$ detection limit, with equivalent width 
\begin{equation}
W_{\rm obs}\geq \frac{(SL)~(\lambda/R)}{(S/N)_\lambda}, \label{eq_wmin}
\end{equation}
where the instrumental resolving power $R=\lambda/\Delta\lambda$ and $SL$ is the significance level in standard deviations.  However, both DS08 and other studies locate \Lya\ lines interactively, which does not follow the strict $W_{\rm min}$ argument used for lines of metal ions.  Determining the significance level of the \Lya\ lines reported by DS08 based {\it a posteriori} on the observed equivalent widths and data S/N, we find $SL>10-15$ for the weakest detections.  Weaker features are certainly visible in the data, but they can reasonably be explained as fixed-pattern noise or other instrumental features.  Indeed, DS08 use $SL\ge10$ when determining the redshift pathlength $\Delta z_{\rm Ly\alpha}$ in their absorber frequencies and cosmological calculations.

\citet{Tripp08} point out that equation~\ref{eq_wmin} is strictly valid only for unresolved features.  Since BLAs are several times wider than the instrumental resolution element for both STIS and \FUSE\ data, equation~\ref{eq_wmin} is even less accurate.  However, we argue that, even if the SL is not rigorously correct, it is still relatively correct, and it gives us a basis for equal comparison between lines.  Broader lines will be shallower for the same equivalent width and thus each pixel will show less contrast from the continuum at full spectral resolution.  Smoothing over the number of pixels equal to the line width (something the human brain does naturally) to first order yields no change in sensitivity for lines of the same equivalent width but different $b$-values. 

Previous BLA studies used the quantity $\log\,(N/b)$ as a detection criterion, reasoning that broader lines require a higher column density (and hence equivalent width) to reach the same line center optical depth.  In particular, \citet{Richter06a} used $\log\,(N/b)>11.3$ [for $N$ in \cd\ and $b$ in \kms] as their detection threshold.  For a constant column density, $\log\,(N/b)$ changes rapidly for narrow lines, but much more slowly at $b>40$~\kms.  For example, $\log\,(N/b)>11.3$ corresponds to $\log\,N_{\rm Ly\alpha}>12.9$ at $b=40$~\kms\ but only 0.2 dex (60\%) higher at $b=65$~\kms.  We argue that setting a detection threshold based purely on equivalent width, $W_{\rm Ly\alpha}\,(S/N)/\Delta \lambda>4$, is roughly equivalent to setting one in $(N/b)$ for BLAs.  It should be noted that $\sim75\%$ of the Possible and Probable BLAs in our sample show $\log\,(N/b)\ge11.3$, and all have $\log\,(N/b)>11.0$.

Total absorption pathlength $\Delta z$ depends on the redshift of the background AGN and the wavelength coverage of the UV spectrograph.  We follow a procedure identical to that discussed in DS08; the local S/N in the data is defined as the mean flux divided by the standard deviation in continuum regions after the data have been smoothed to the resolution element.  The $S/N(\lambda)$ is then modified by setting regions with strong IGM, Galactic, and intrinsic absorption systems and instrumental features equal to zero.  $W_{\rm min,Ly\alpha}(z)$ is calculated from the $S/N(\lambda)$ vector.  The pathlength $\Delta z(W_{\rm min})$ is then the sum of pixels where $W>W_{\rm min}$.  As in DS08, we omit regions within 500 \kms\ of the Galaxy and within 1500 \kms\ of the AGN to eliminate (most) absorbers intrinsic to either the AGN or the Local Group.  Cosmologically corrected pathlength $\Delta X$ is calculated in an entirely analogous manner, using $dX\equiv(1+z)^2\,\bigl[\Omega_m(1+z)^3+\Omega_\Lambda\bigr]^{-1/2} dz$.  Throughout this paper we assume a flat ($\Omega_m$, $\Lambda$) cosmology with $H_0=(70\rm~km~s^{-1}~Mpc^{-1})\,h_{70}$, $\Omega_m=0.261$, $\Omega_\Lambda=0.716$, and $\Omega_b=0.0455\,h_{70}^{-2}$ \citep{Spergel07}.  The maximum path lengths available in each sight line for absorbers of any strength are listed in Table~1.  The total pathlength surveyed in all seven sight lines is $\Delta z_{\rm tot}=2.193$ ($\Delta X_{\rm tot}=1.773$).

For the numerator of $(d{\cal N}/dz)_{\rm BLA}$, we have several options to choose from, depending how much faith we place in our BLA designations.  The most skeptical view accepts only our Probable sample, which with one-sided Poisson statistics gives ${\cal N}_{\rm BLA}=15^{+5}_{-4}$ and $d{\cal N}/dz=7\pm2$.  A more inclusive view includes both the Probable and Possible groups: ${\cal N}_{\rm BLA}=63^{+9}_{-8}$ and $d{\cal N}/dz=29\pm4$.  Since $\sim50$\% of the BLAs survive the statistical correction process described in Section~3.2, we adopt instead an intermediate census: all of the Probable BLAs and 50\% of the Possible sample: ${\cal N}_{\rm BLA}=39^{+7}_{-6}$.  Given the uncertainties surrounding BLA identification, we believe pure Poisson uncertainties are far too optimistic.  We adopt the skeptical and inclusive censuses as our lower and upper bounds: ${\cal N}_{\rm BLA}=39\pm24$, $(d{\cal N}/dz)_{\rm BLA}=18\pm11$ or, in comoving coordinates, $(d{\cal N}/dX)_{BLA}=22\pm14$.  Despite the corrections above, uncertainties in pathlength are small ($\la10$\%), so we ignore errors in the denominator.

\subsection{Overlap between BLAs and Metal-Line Absorbers}

Because both BLAs and highly-ionized metal lines (\OVI, \NV, \NeVIII, etc.) are thought to trace WHIM gas, it is instructive to look at the overlap in these two samples.  \OVI\ has by far the best detection statistics of any FUV intergalactic metal line, with $\sim100$ detections in the low-$z$ IGM surveys \citep{DS08,Tripp08,ThomChen08a}.  In \citet{Stocke07}, we used $\log\,N_{\rm OVI}\geq13.2$ and $\log\,N_{\rm OVI}<13.2$ as the threshhold between a good \OVI\ detection and a reliable non-detection based on the distributions of observed detections and $4\sigma$ upper limits.  Using the same threshold, we find that four probable BLAs show \OVI\ absorption, while six show \OVI\ non-detections; a detection rate of $\sim40$\%.  For the larger sample of probable-plus-possible BLAs, the numbers rise to 8 detections and 33 non-detections for an \OVI\ detection rate of $\sim20$\%.  Using the same criteria, the large DS08 survey ($\sim650$ \HI\ systems) features 69 \OVI\ detections and 293 non-detections (19\%).  However, this sample mixes broad with narrow \HI\ lines.  If we instead define a control sample of non-BLAs as all DS08 \HI\ systems with $b_{\rm HI}<40$ \kms\ (516 systems), there are 47 \OVI\ detections and 245 non-detections (16\% detection rate, half that of our Probable BLA sample).  The \OVI\ detection rate is slightly lower (14\%) if we use the DS08 $\bLW$ measurements to define the non-BLA control group. 

The coincidence of probable BLAs and \OVI\ detections (40\%) is several times higher than in the larger, narrow \Lya\ absorber sample ($\sim15$\%).  This suggests that BLAs and \OVI\ are tracing the same material, but the small number of systems in the BLA sample reduces the significance.  It is worth noting that the 40\% \OVI\ detection rate is a bit higher than the $\sim25$\% fraction of \HI\ systems that show metal absorption in {\it any} ion reported by DS08.  This suggests that BLAs are accurately tracing WHIM irrespective of metal enrichment.  Unfortunately, the detection statistics for other ions are too poor to draw any conclusions.

\subsection{The Galaxy-BLA Connection}

One of the strongest potential advantages of detecting the WHIM using BLAs is that these absorbers are not affected by the metallicity of the gas, so that even metal-free WHIM can be detected.  We would expect these low-metallicity and metal-free absorbers to be found in regions far from galaxies, unlike the \OVI\ absorbers which are typically found within $\sim800$~kpc of the nearest L$^*$ galaxy \citep{Stocke06,WakkerSavage09} and even closer to sub-L$^*$ galaxies \citep{Stocke06}.  Unfortunately, only eight of the BLAs reported here are found in sky regions surveyed for galaxy redshifts complete to L$^*$ or below; these are listed in Table~3 in increasing order of galaxy separation.  Dividing this sample into \OVI\ detections and non-detections at a consistent level of log\,$N_{\rm OVI}=13.2$ (DS08), we find nearest galaxies at $0.75-2.9~h^{-1}_{70}$~Mpc for the \OVI\ non-detections and $0.06-0.49~h^{-1}_{70}$~Mpc for the detctions.  Therefore, we find some evidence that BLAs are tracing WHIM gas more remote from galaxies than by using \OVI\ absorption as a WHIM tracer.

\begin{deluxetable}{llcccl}
  \tabletypesize{\footnotesize}
  \tablecolumns{6} 
  \tablewidth{0pt} 
  \tablecaption{Galaxy-BLA Relationship in Well-Surveyed Fields}
  \label{tab_nearestgal}
  \tablehead{\colhead{AGN}   &
           \colhead{$z_{\rm abs}$}         &
	   \colhead{$b_{\rm HI}$\tnma}        & 
	   \colhead{BLA}&
           \colhead{log\,\NOVI\tnmb} & 
           \colhead{d} \\
	   \colhead{}&
	   \colhead{}&
	   \colhead{(\kms)}&
	   \colhead{class}&
	   \colhead{(\cd)}&
	   \colhead{(Mpc)}  }
\startdata 
 PG\,1259$+$593  &0.00229 &  44 COG &A&$  13.7$:\tnmc & 0.06 \\
 PKS\,0405$-$123 &0.16678 &  75: LW &B&$  14.1\pm0.2$ & 0.11 \\
 PKS\,0405$-$123 &0.09659 &  70: LW &B&$  13.7\pm0.2$ & 0.27 \\ 
 PKS\,0405$-$123 &0.08139 &  53 LW  &A&$  13.3\pm0.2$ & 0.49 \\
 PG\,1116$+$215  &0.06072 &  54 LW  &B&$ <13.07    $  & 0.75 \\
 PG\,1116$+$215  &0.09279 & 100: LW &A&$ <13.07 $     & 1.3  \\
 PG\,1116$+$215  &0.01635 &  51 LW  &B&$ <13.05    $  & 2.0  \\
 PG\,1116$+$215  &0.08587 &  53 LW  &B&$ <13.04    $  & 2.9  \\
\enddata
\tablenotetext{a}{Consensus $b$ value from Table~2}
\tablenotetext{b}{O\,VI column density from DS08}
\tablenotetext{c}{Detection from \citet{Richter04} based on nighttime-only \FUSE\ data.  Formal significance level is low ($<3\sigma$), however absorption appears over an exceptionally broad velocity range ($-110<v<+220$ \kms).
}
\end{deluxetable}

There is one slightly controversial absorber that we have counted as an \OVI\ detection in the above accounting: the $z=0.00229$ \Lya\ absorber toward PG\,1259$+$593 is 60~kpc from the $14^{th}$-mag edge-on late-type spiral galaxy UGC\,8146.  DS08 report this absorber as an \OVI\ non-detection according to their $4\sigma$ detection threshhold, but \citet{Richter04} report a low-significance, very broad \OVI\ detection at $N_{\rm OVI}=5\times10^{13}$~\cd\ based on night-only \FUSE\ data.  We thus include this as an \OVI\ detection.

\section{BLA Cosmology}

Only about half of the baryons can be accounted for in the local universe.  The \Lya\ forest makes up only about 30\% of the total predicted baryons at $z\approx0$ \citep{Penton04,Lehner07,DS08} while collapsed structures (stars, galaxies, etc.) make up another $\sim10\%$ \citep{SalucciPersic99}.  Much of the remainder is expected to lie in the ionized phases of the IGM above $10^5$~K.  The $\sim100$ \OVI\ absorbers observed at $z<0.4$ have been used to trace the WHIM phase and can account for an additional $\la10$\% of the baryons (DS08), but this estimate relies on metallicity and ionization-fraction assumptions that make the quantity uncertain \citep{Danforth09}.  The \OVI\ WHIM surveys require metal enrichment, leaving open the possibility that a significant population of metal-poor IGM clouds may contribute to the baryon census.  The strength of BLA surveys is their ability to trace gas at $T=10^5-10^6$~K independent of chemical enrichment.  While there is some overlap with \OVI\ WHIM absorbers, BLAs open a new window on the cosmic baryon census.

\subsection{Baryon Fraction Traced by Broad \HI}

The mass fraction of the local universe traced by broad \Lya\ absorbers can be determined by dividing the total hydrogen column density by the total observed pathlength
\begin{equation}
\Omega_{\rm BLA}=\biggl(\frac{\mu\,m_H\,H_0}{c\,\rho_{\rm crit}}\biggr)\,\frac{\sum N_H}{\sum \Delta X}~. 
\label{eq_omegabla1}
\end{equation}
Since the vast majority of IGM hydrogen is ionized, total hydrogen column density can be approximated for any given absorber by $N_{\rm H}\approx N_{\rm HI}/f_{\rm HI}(T)$.  Neutral hydrogen column \NHI\ can be measured directly in most cases.  However, the hydrogen neutral fraction $f_{\rm HI}$ is determined by both photoionization from the metagalactic radiation field and ionization due to (thermal) electron collisions.  At $z\approx0$, the ionizing background produces an \HI\ photoionization rate $\Gamma_H=\rm 3.2^{+2.0}_{-1.2}\times10^{-14}~s^{-1}$ as derived in \citet{Shull99} from populations and radiative transer calculations of Seyferts, QSOs, and starbursts.  For electron impact, the hydrogen ionization rate can be approximated as 
\begin{equation}
  C_i(T)=(5.83\times10^{-11}~{\rm cm^3\,s^{-1}})\,\frac{T^{1/2}~\exp[-1.58\times10^5/T]}{1 + 1.58\times10^6/T}, 
\end{equation}
where $1.58\times10^5~K=(I_H/k_B)=(13.6~eV)/k_B$.  The critical density where collisional ionization equals photoionization is then $(n_e)_{\rm crit}=\Gamma_H/C_i(T)$.  For borderline WHIM temperatures ($T=10^5$~K), the ionization rate is $C_i=3.6\times10^{-9}~\rm cm^3~s^{-1}$ and the critical density is $(n_e)_{\rm crit}=8.9\times10^{-6}\rm~cm^{-3}$.  However, collisional ionization becomes more and more dominant at higher temperatures, and at $\log\,T=5.5~(6.0)$, $C_i(T)=1.7(3.1)\times10^{-8}\rm~cm^3~s^{-1}$ and $(n_e)_{\rm crit}=1.9(1.0)\times10^{-6}\rm~cm^{-3}$ corresponding to overdensities of $\delta<10$ at $z\sim0$.  

Photoionization is thus an important consideration at low temperatures ($k\,T\ll1$~Ryd), but at temperatures near or above the \OVI\ peak in CIE ($\log\,T=5.5-6.0$), we have $(n\rm _e)_{crit}=few\times10^{-6}~cm^{-3}$.  At those low densities, in photoionization equilibrium, $f_{\rm HI}=n_e\,(\alpha_H)/(\Gamma_H)\sim(1.3\times10^{-5})~[n_e/10^{-6}\rm~cm^{-3}]$ we adopt $\Gamma_H = 3.2\times10^{-14}~\rm s^{-1}$ and $\alpha_H=4.1\times10^{-13}\rm~cm^3~s^{-1}$ (case-A at $10^4$~K).  If $n_e\approx(n_e)_{\rm crit}$, collisional ionization could double the ionization rate and halve $f_{HI}$ in the formula above ($0.65\times10^{-5}$).  A BLA with $n_e=(n_e)_{\rm crit}$ and $N_{HI}=3\times10^{13}~\rm cm^{-2}$ would have $N_H\sim5\times10^{18}$~\cd\ and line-of-sight dimension $L_{\rm BLA}=N_H/n_H\sim5\times10^{24}~\rm cm=1.5~Mpc$.  An unvirialized cloud this size would exhibit a 100 \kms\ broadening of its \Lya\ linewidth due to Hubble flow from one side to another.

We adopt $f_{\rm HI}(T)$ values derived from a set of CLOUDY simulations (solid curve in Fig.~\ref{fig_fhi}) featuring both collisional and photoionization ($\log\,U=-2$, typical of the low-$z$ IGM) as the most valid approximation to neutral fraction.  At WHIM temperatures, this closely follows the CIE neutral fraction, but diverges quickly at $T<10^5$~K.  The model ionization parameter $U=n_\gamma/n_H=0.01$ is typical of the low-$z$ IGM, but large changes in the model ionizing field will produce only small changes in $f_{\rm HI}(T)$ at WHIM temperatures.  Figure~{\ref{fig_fhi} shows model curves for $f_{\rm HI}(T,U)$ based on photo-thermal CLOUDY models with $\log\,U=-2$ and $\log\,U=-1$; they differ by $\sim0.3$ dex at $T\approx10^5$~K, but by only $\sim0.1$ dex at $T\ga10^{5.5}$~K.

Path length $\Delta X(N_{\rm HI})$ for each sight line is calculated as described in Section~3.1.  Of the 63 Probable and Possible BLA candidates, most are strong enough that $\Delta X(N_{\rm HI})\approx \Delta X_{\rm max}$.  However the survey of the weaker BLAs is only $\sim80\%$ complete.  We correct for completeness by dividing column density \NHI\ by the corresponding completeness in their respective data (all correction factors were between 0.8 and 1.0).  The BLA mass fraction is calculated by modifying equation~\ref{eq_omegabla1} to
\begin{equation}
\Omega_{\rm BLA}=\biggl(\frac{\mu\,m_H\,H_0}{c\,\rho_{\rm crit}}\biggr)~\frac{\sum_{i,j} \bigl(f_{HI}^{-1}\,N_{\rm HI}\bigr)_{i,j}\,\bigl(\frac{\Delta X_{{\rm max},j}}{\Delta X_{i,j}}\bigr)}{\sum_j \Delta X_{{\rm max},j}}~.\label{eq_omegabla2}
\end{equation}


\begin{figure}
  \epsscale{1.2}\plotone{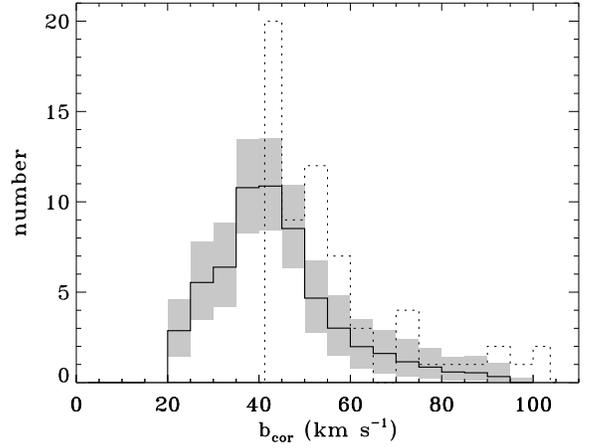} 
  \caption{Distribution of input and corrected $b$-values.  Input
  sample (dotted) is corrected based on randomly-selected $b_{\rm
  COG}/b_{\rm LW}$ ratios from DS08.  Solid line and shaded boxes show
  the median and $\pm1\sigma$ distribution of $10^4$ Monte-Carlo
  simulations per absorber.  Any BLA candidate corrected to $b<40$
  ~\kms\ is dropped from the calculation of $\Omega_{\rm BLA}$
  performed later in this paper.}\label{fig_bdist}
\end{figure}

\begin{figure}
  \epsscale{1.2}\plotone{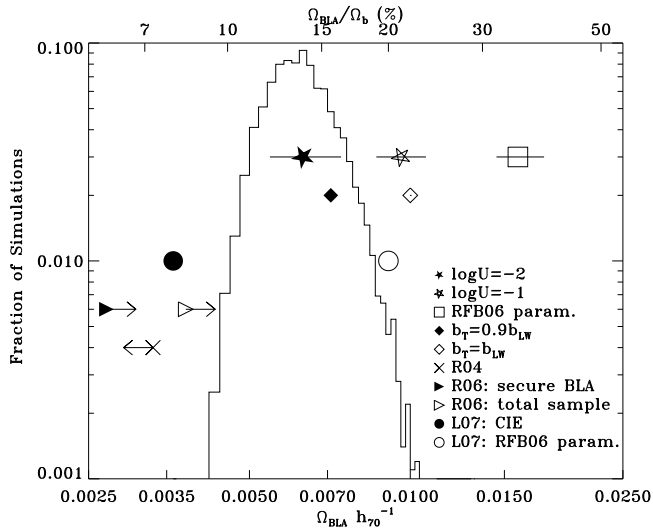} 
  \caption{Distribution of $\Omega_{\rm BLA}$ values from $10^4$
  Monte-Carlo simulations (histogram) with median value (filled star)
  and $\pm1\sigma$ range (horizontal line) $\Omega_{\rm
  BLA}=6.3^{+1.1}_{-0.8}\times10^{-3} \,h_{70}^{-1}$.  Various other
  assumptions are also shown for comparison.  Using neutral fractions
  based on a CLOUDY model with $\log\,U=-1$ (rather than $\log\,U=-2$)
  shifts the simulated $\Omega_{\rm BLA}$ distribution to larger
  values $\Omega_{\rm BLA}=(9.6^{+1.1}_{-1.0})\times10^{-3}$ (open
  star).  Using the \citet{Richter06b} $f_{\rm HI}$ parametrization
  yields $\Omega_{\rm BLA}=(15.8^{+1.9}_{-1.4})\times10^{-3}$ (open
  square).  Assuming $\bLW=\btherm$ gives a solution $\sim2\sigma$
  higher than the simulations (open diamond).  The \citet{Richter06b}
  assumption that $\btherm=0.9\times\bLW$ (open diamond) gives nearly
  the same value we get from our statistical treatment.  Literature
  $\Omega_{\rm BLA}$ values and upper limits from \citet{Richter04},
  \citet{Richter06a}, and L07 are shown for comparison.  The
  equivalent baryon fraction, $\Omega_{\rm BLA}/\Omega_b$, is plotted
  on the top axis for comparison.} \label{fig_omegadist}
\end{figure}

Given the uncertain relationship between $b_{\rm obs}$ and $\btherm$ for any given absorber, we apply a statistical procedure to better ascertain thermal line widths.  First, we assume that $\bCOG=\btherm$ where available.  For the remainder, we perform a Monte-Carlo simulation, correcting each observed Probable and Possible BLA linewidth as follows.  Each candidate is ``corrected'' using a randomly selected $\bCOG/\bLW$ ratio from the 138 absorbers in DS08 with well-determined $\bCOG$ (Fig.~\ref{fig_DS08}).  Each BLA candidate is simulated $10^4$ times and the resulting distribution of $b_{\rm cor}$ is shown in shown in Fig.~\ref{fig_bdist}.  The survival fraction of an individual BLA candidate (i.e., $b_{\rm cor}\ge40$~\kms) is ($62\pm5$)\% although this includes seven $\bCOG\geq40$ \kms\ absorbers which are not corrected.  Roughly 50\% of the non-COG BLA candidates survive the correction process.


The corrected linewidths are used to calculate $\Omega_{\rm BLA}$ for each Monte-Carlo simulation.  The gas temperature is assumed to be $T=60\,b_{\rm cor}^2$~K ($b$ in \kms).  Neutral fraction as a function of temperature $f_{\rm HI}(T)$ is determined from a CLOUDY simulation as discussed above, and the total hydrogen column density is calculated $N_{\rm H}=N_{\rm HI}/f_{\rm HI}$ for each absorber.  Summing over all BLAs, we derive $\Omega_{\rm BLA}$ according to equation~(\ref{eq_omegabla2}).  The full $\Omega_{\rm BLA}$ distribution for $10^4$ simulations (Fig.~\ref{fig_omegadist}) shows a median and $\pm1\sigma$ value of of $\Omega_{\rm BLA}=6.3^{+1.1}_{-0.8}\times10^{-3}\,h_{70}^{-1}$ or a baryon fraction of $\Omega_{\rm BLA}/\Omega_b=14^{+3}_{-2}$\%.  Varying some of these assumptions changes $\Omega_{\rm BLA}$ as discussed below.

\subsubsection{Systematic Uncertainties and Trends in $\Omega_{\rm BLA}$}

There are a number of uncertainties and assumptions present in our $\Omega_{\rm BLA}$ determination which can affect $\Omega_{\rm BLA}$ in several ways.  Since we are inferring a total hydrogen column density from an observed neutral trace component, both $b$-values and \NHI\ have a very large lever arm with which to act on the total baryon count. 

First, we assume that the COG-determined (or statistically corrected) line width is entirely due to thermal broadening.  As discussed above, even a noise-free COG can overestimate a true line width by $20-50$\% in the case of undetected, blended components.  If we assume all $\bCOG$ values overestimate $\btherm$ by 20\% (as discussed in Section~2.1), the post-correction survival fraction of BLAs becomes much lower ($34\pm4$\%) and $\Omega_{\rm BLA}=(2.4^{+0.6}_{-0.4})\times10^{-3}$, more than a factor of two lower than our assumed $\bCOG=\btherm$ value.

Second, since IGM absorbers are thought to be quite large in extent and generally unvirialized, there is potentially a Hubble expansion between one side of the cloud and another.  Using a typical absorber scale of 350 kpc \citep{Paper2,DS08}, we might expect a differential $\Delta v\sim 24\,h_{70}$ \kms\ between the two sides.  This would add in quadrature with the other non-thermal line-broadening effects.  Running the simulation with this additional non-thermal correction in place yields a lower mass fraction from BLAs: $\Omega_{\rm BLA}=3.7^{+1.0}_{-0.7}\times10^{-3}$.

Third, it is unknown how many of our perceived broad lines have a narrow component to them, and conversely, how many blended \Lya\ forest lines contain a weak, broad component dominated by a narrow absorber.  The BLA column density is overestimated for the former lines and uncounted for the latter.  Future observations with high S/N ($\ga30$) from the Cosmic Origins Spectrograph may disentangle some of these lines, but many blended, multiphase systems will likely remain permanently entangled (see Fig.~3), leaving the final BLA baryon census uncertain.  

We assume that the detectability of a line of a particular width/depth is a function purely of the local signal-to-noise ratio of the data.  Based on this assumption, the data are not less than 80\% complete for the range of BLA candidates.  If this assumption is optimistic and the data is actually incomplete to $\sim50\%$ in some cases, the numerator in Eq.~\ref{eq_omegabla2} may rise by as much as $\sim60\%$ for {\it some} absorbers in the sample.  The contribution to $\Omega_{\rm BLA}$ from these terms would increase accordingly, but we expect the correction to the {\it total} $\Omega_{\rm BLA}$ sum to be minor.

In this work, we have considered only absorbers with $\bLW\ge40$~\kms.  However, the distribution of $\bLW/\bCOG$ values (Figure~2) shows that 12\% (20/164 COG solutions) have $\bLW/\bCOG<1$.  Correcting narrow lines by ratios less than unity will result in a statistical line {\it broadening} and could, in principle, create additional BLA candidates.  A closer look at the data suggests that ``scatter-up'' is a small effect.  The smallest $b$-ratio in the DS08 distribution is $\bLW/\bCOG\sim0.7$, so only absorbers with $28\le\bLW<40$ \kms\ would be broadened to $b_{\rm cor}\ge40$ \kms.  There are 82 \Lya\ absorbers in DS08 in the seven sight lines examined here.  On average, a maximum of $82\times0.12\approx10$ absorbers would be scattered up.  However, due to the opposing trends in $\bLW$ and $\bLW/\bCOG$ distributions, the likely number is smaller ($4-5$).  Additionally, we have not scrutinized the full sample of $30\le b_{\rm LW}<40$ \kms\ absorbers for obvious multiple component structure, which can only reduce the likely number of additional BLAs.  Thus, we expect these ``scattered-up'' narrow lines to be a small correction to $\Omega_{\rm BLA}$.

\subsection{Comparison to Previous Work}

Previous BLA studies have approached the issues of line identification, line width, and $f_{\rm HI}$ in different ways and most have used some or all of the same datasets studied here.  It is instructive to look at the assumptions in and results from these studies to see how different systematics affect \dndz\ and $\Omega_{\rm BLA}$.  All of these values are shown in Fig.~\ref{fig_omegadist} in comparison to our results. 

\citet{Richter04} study $\sim9$ BLAs in the PG\,1259$+$593 sight line.  They find $(d{\cal N}/dz)_{\rm BLA}\approx23$ and calculate $f_{\rm HI}$ in CIE from inferred temperature.  They note that $\bLW$ overestimates $\btherm$ and thus quote only an upper limit $\Omega_{\rm BLA}\leq3.3\times10^{-3}\,h_{70}^{-1}$, roughly half of our result.  However, this is a single sight line and the difference may be explainable by cosmic variance.

\begin{figure*}
  \epsscale{1}\plotone{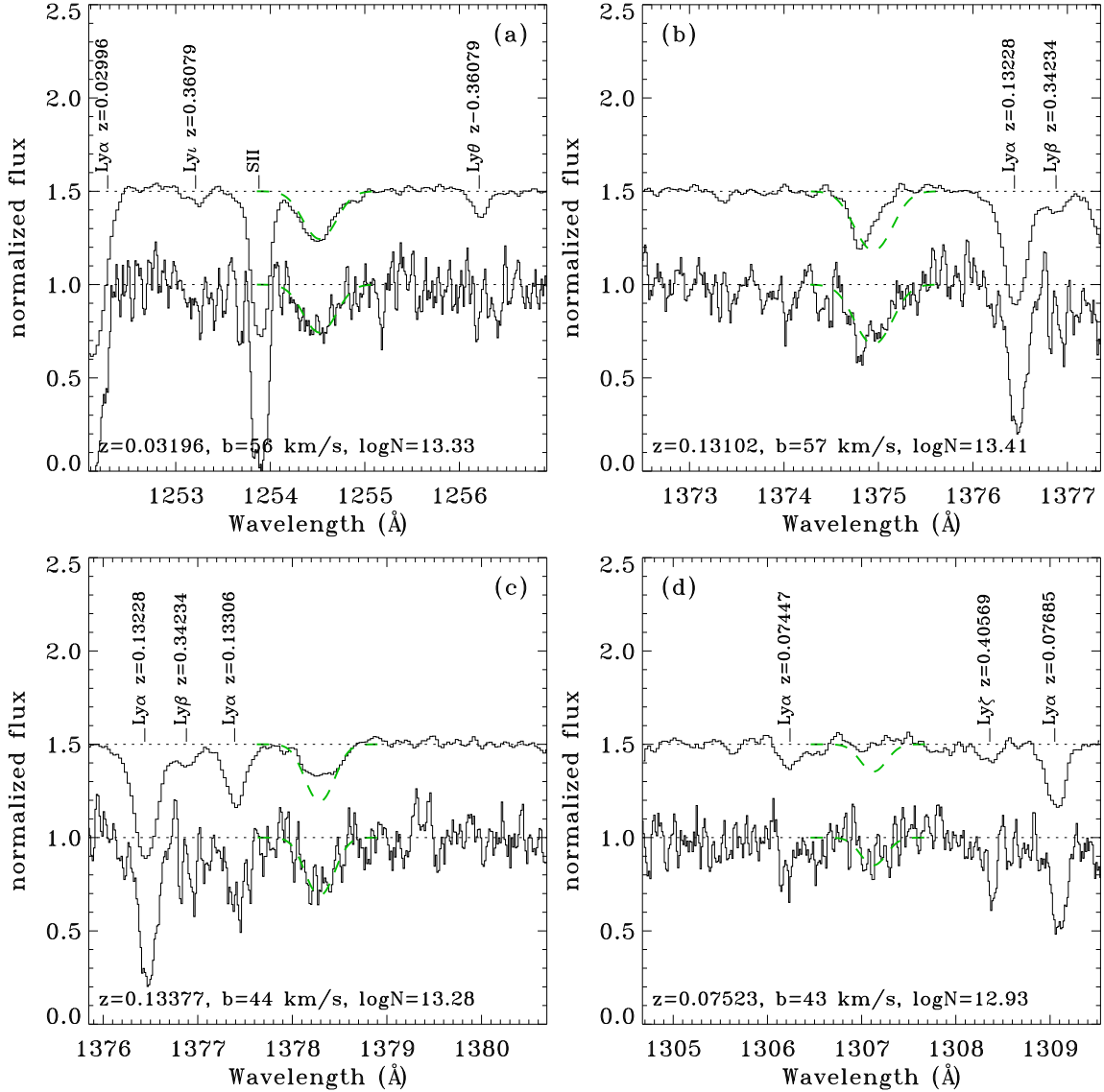} 
  \caption{Examples of four Possible BLAs toward PKS\,0405$-$123
  measured in STIS/E140M data which can either be confirmed or refuted
  based on $S/N\approx50$ COS/G130M Early Release Observations.  The
  bottom spectra in each panel show the STIS/E140M data used in our
  analysis to derive consensus $N$ and $b$ values (listed).  The upper
  spectra are COS data with prominent lines labeled and offset upward
  by 1.5.  Panel (a) shows a BLA candidate which has essentially the
  same profile in both datasets.  Panels (b) and (c) show absorbers
  which show a significantly different structure in the high-S/N COS
  data.  Panel (d) shows a reported Possible BLA which does not appear
  in the COS spectrum.  All data are smoothed to their respective
  resolution elements.  The consensus line fit of the STIS data is
  marked by a dashed line in both profiles.} \label{fig_cos}
\end{figure*}

\citet{Richter06a} analyze H\,1821$+$643 and PG\,0953$+$415 and bring in the results from PG\,1259$+$593 \citep{Richter04} and PG\,1116$+$215 \citep{Sembach04}.  They find 20 and 49 BLAs in their ``secure''  and total samples, with $(d{\cal N}/dz)_{\rm BLA}=22$ and $(d{\cal N}/dz)_{\rm BLA}=53$, respectively.  Assuming CIE, they find $\Omega_{\rm BLA}>2.7\times10^{-3}\,(3.8\times10^{-3})\,h_{70}^{-1}$, but the value becomes unphysically large ($>\Omega_b$) if CIE $+$ photoionization is assumed (dashed curve in Figure~1).

\citet{Lehner07} compile data from seven sight lines including four from previous work (PG\,1259$+$593 \citep{Richter04}, HE\,0226$-$4110 \citep{Lehner06}, HS\,0624$+$6907 \citep{Aracil06b} and PG\,1116$+$215 \citep{Sembach04}), two from as-yet-unpublished private communication (PG\,0953$+$415, Tripp \etal\ in prep. and H\,1821$+$643, Sembach \etal\ in prep), a detailed reanalysis of PKS\,0405$-$123.  They perform quality and column-density cuts on 341 \Lya\ lines and end up with $\sim60$ BLAs for a density $(d{\cal N}/dz)_{\rm BLA}=30\pm4$.  To deal with blended components and non-thermal broadening, L07 randomly eliminate 1/3 of their BLAs with $40<b<65$ \kms\ (all absorbers with $b>65$~\kms\ are kept).  Nonthermal broadening is assumed to be $\sim10$\% of the total line width \citep{Richter06b}, so temperatures are calculated assuming $b_{\rm T}\approx 0.9\,b_{\rm obs}$.  They calculate $\Omega_{\rm BLA}$ assuming both CIE and Richter's CIE $+$ photoionization parametrization and find $\Omega_{\rm BLA}=3.6\times10^{-3}\,h_{70}^{-1}$ and $9.1\times10^{-3}\,h_{70}^{-1}$, respectively or $\Omega_{\rm BLA}/\Omega_b=8\%$ and $20\%$, for each case.

At higher redshift, \citet{Prause07} observed BLAs in the spectra of five $1.34<z<1.94$ AGN using both the near-UV STIS/E230M grating on HST and the ground-based UV Echelle Spectrograph at the ESO Very Large Telescope.  In total, they found 9 good BLA candidates and an additional 29 tentative cases in the redshift range $0.9\la z \la 1.9$.  They derive a value of $\Omega_{\rm BLA}=(2.2\pm0.1)\times10^{-3}\,h_{70}^{-1}$ for the 9 good BLA candidates and $\Omega_{\rm BLA}=(14\pm2)\times10^{-3}\,h_{70}^{-1}$ for the entire sample.

\subsubsection{Broad Absorbers in Cosmic Origins Spectrograph Observations}

Late in the analysis process, we obtained observations of PKS\,0405$-$123 by the HST/Cosmic Origins Spectrograph \citep{Green10,Osterman10} as part of its public Early Release Observations (ERO) program.  These data were obtained in the G130M grating ($\rm1132~\AA<\lambda<1468~\AA$) with a nominal resolution $R\sim20,000$ ($\Delta v=15$ \kms).  Seven orbits in total ($\sim17$~ksec) were devoted to ERO observations, with three taken prior to when an accurate focal alignment was achieved and four afterward.  A close examination of the data shows no difference in line profiles for any of the ISM or IGM lines of interest, so observations from all seven orbits were aligned and coadded.  

The resulting spectrum is of exquisite quality with $S/N\ge40$ per nominal seven-pixel resolution element at most locations.  Owing to the differences in detector technology and the different grating positions used in the observations, the COS data are free of much of the fixed-pattern noise that plagues the corresponding STIS/E140M observations.  This, coupled with the very high S/N, makes COS ideal for verifying the BLA candidates discussed in this paper.

Eighteen of the BLA candidates toward PKS\,0405$-$123 measured in STIS/E140M data are also covered in the COS observations.  Five of these cases are blended components of strong absorption systems, which are difficult to confirm or refute, but the COS data is consistent with the STIS observations.  Of the 13 weaker absorption lines, seven show a good match between STIS and COS data.  However the STIS-measured line profile is substantially different in four cases and missing altogether in another three.  Figure~\ref{fig_cos} shows several examples where BLA candidates can be either confirmed or refuted based on COS observations.

This is a good demonstration of the importance of fully understanding the instrumental effects of a particular spectrograph.  While the COS data are not free of fixed pattern noise, it is independent of that in the STIS observations.  Furthermore, it is evidence that the $\sim10\times$ sensitivity increase of COS over STIS will revolutionize the study broad \Lya\ absorbers, both through higher S/N and sensitivity to numerous, fainter targets.  Since this work relies on a uniform analysis of consistent datasets by independent groups, we do not change any of our BLA designations in light of new, higher-quality data from COS.  However, we note the results of our COS absorber verification where possible in the individual absorber comments in the Appendix.

\section{Conclusions and Summary}

Broad \Lya\ absorbers are a potentially powerful method of measuring the extent and distribution of gas at $T\geq10^5$~K in the intergalactic medium without relying on metal enrichment.  The small hydrogen neutral fraction even at WHIM temperatures  will result in broad, shallow \HI\ profiles that can be translated into temperatures and total hydrogen column densities.  Unfortunately, BLAs present some observational challenges and ambiguities.  Broad, shallow absorbers are difficult to detect in data of only moderate S/N and in multi-phase systems.  Detected BLAs are strongly biased toward cooler temperatures where lines are relatively narrower and the neutral fraction is higher.  Furthermore, identification of {\it bona fide} BLAs relies crucially on the line component structure, knowledge of instrumental features and an accurate continuum definition. 

We attempt to work around these problems by independent reduction and analysis of seven AGN sight lines containing 119 purported broad \Lya\ lines reported in the literature (mainly DS08 and L07 and sources therein).  We assign consensus values for column density and line width based on two independent analyses by two different research groups.  The purported BLAs are split into three qualitative categories based on the consensus linewidth, detailed analysis of the absorption profile, and other factors.  Probable BLAs (15 systems) are those showing $\bLW>60$ \kms\ and no obvious asymmetry or component structure.  Systems with curve-of-growth determined linewidths $\bCOG>40$ \kms\ are also deemed Probable BLAs.  Possible BLAs (48 systems) are those with $40<\bLW<60$ \kms, those absorbers with potential component structure or asymmetries, or some plausible reason to doubt their identity as BLAs.  The remaining systems (56) are ruled out as being BLAs for a number of reasons: alternate line identification, $\bLW<40$~\kms, probable component structure, or simply failing to appear in our reduction of the data.

Taking all of the probable BLAs and $50\pm50$\% of the possible category, we see $39\pm24$ BLAs along a total redshift pathlength $\Delta z=2.193$ ($\Delta X_{\rm tot}=1.773$) in the seven AGN sight lines surveyed.  This gives a detection frequency of $d{\cal N}/dz_{\rm BLA}=18\pm11$ ($d{\cal N}/dX_{\rm BLA}=22\pm14$).  This frequency is similar to that of \OVI, another potential WHIM tracer with $(d{\cal N}/dz)_{\rm OVI}=15\pm3$ \citep{DS08,Tripp08}, though the BLA frequency has considerably greater uncertainty.  Indeed, while the detection or non-detection of highly ionized ions (\OVI, \NV, \NeVIII, etc) was not taken into account in our BLA categorization, 40\% of the probable BLAs and 20\% of the combined probable-plus-possible samples show reasonable \OVI\ detections.  The incidence of \OVI\ detections in narrow \Lya\ lines is $\sim15$\%.

The relationship of WHIM to galaxies is another key area of interest.  It is likely that small galaxies with weak gravitational fields are important for IGM heating and enrichment \citep[e.g.,][]{Stocke04}.  Unfortunately, surveys for low-luminosity galaxies tend to be unreliable at redshifts greater than a few hundredths.  However, we found eight BLAs in regions of fairly complete galaxy surveys ($L\ga0.2\,L^*$) that provide nearest-galaxy distances of 60 kpc out to nearly 3 Mpc.  The detection of \OVI\ in conjunction with broad \Lya\ was highly correlated with galaxy distance as the four BLAs with $d\la600$ kpc showed \OVI\ detections at some level, while the four BLAs at $d\ga600$ kpc appear to be free of \OVI\ absorption.  This result gives significant support for BLAs probing gas that \OVI\ surveys do not detect.  Using the above \OVI\ detection statistics, our BLAs suggest that $\sim80$\% of the baryons are not already accounted for in \OVI\ surveys.

A main and crucial uncertainty in BLA surveys is disentangling thermal and non-thermal line broadening, as measured line width overpredicts the thermal linewidth, often by an unknown amount.  Previous studies have recognized this phenomenon and dealt with it in a variety of ways, including arbitrarily throwing out some fraction of measured BLAs from a sample or scaling measured line widths by some uniform factor.  We approach the problem by looking at curve-of-growth $b$-values ($\bCOG$): the doppler b-parameter measured from a single line (typically \Lya) overpredicts that from a full COG analysis \citep{Paper2} by a factor of $\sim1.5\pm0.9$ (though we also show how even a COG can overpredict the true linewidth).  Since most broad \Lya\ systems are fairly weak, confirmation in higher-order Lyman lines is usually not possible.  Instead, we statistically correct for the single-line $b$-value overprediction based on the observed $\bLW/\bCOG$ distribution in DS08.  We find in our simulations that $55\pm5$\% of the reported BLAs survive the correction process with $b_{\rm cor}\ge40$ \kms, corresponding to $T\geq10^5$~K.  

From temperatures derived from line widths, we can estimate the neutral fraction of a particular absorber and hence the total hydrogen column associated with a particular \HI\ detection.  The total amount of gas at $T>10^5$~K traced by these BLAs can then be estimated as a fraction of the closure density of the Universe.  We use a Monte-Carlo simulation to statistically correct the observed BLA linewidths.  Our median value of $\Omega_{\rm BLA}=6.3^{+1.1}_{-0.8}\times10^{-3}\,h_{70}^{-1}$ ($\Omega_{\rm BLA}/\Omega_b=14^{+3}_{-2}$\%) based on $10^4$ Monte-Carlo simulations of each BLA.  Since $\sim80$\% of these BLA baryons are not in \OVI\ systems, the combination of \OVI\ $+$ BLA WHIM searches can account for $\Omega_{\rm WHIM}/\Omega_b\sim20\%$.  This includes $\sim8$\% from \OVI\ (DS08) and an additional $\sim12$\% in metal-poor BLAs.  It is clear that systematic uncertainties involved in BLA surveys are comparable to or larger than the statistical fluctuations from cosmic variance among the sight lines, and more work must be done to understand the individual systems.  

To illustrate the importance of methodology and individual systems to the baryon census, we compare our result to that of L07.  L07 used the same set of AGN sight lines used here and a more inclusive set of BLA candidates.  They assume $b_{\rm T}=0.9b_{\rm LW}$ as in \citet{Richter06b}, calculated hydrogen neutral fraction via a CIE assumption (very similar to the CLOUDY $\log\,N=-2$ model used here), and randomly eliminated 1/3 of the $40<b<65$ \kms\ absorbers which gave a result of $\Omega_{\rm BLA}=3.6\times10^{-3}\,h_{70}^{-1}$ or $\Omega_{\rm BLA}/\Omega_b=8\%$ or $\sim60$\% of our value.  If, instead, $f_{\rm HI}$ is based on the CIE $+$ photoionization model of \citet{Richter06b}, the L07 value rises to $\Omega_{\rm BLA}=9.1\times10^{-3}\,h_{70}^{-1}$, or $\sim40$\% larger than our result (see Fig~\ref{fig_omegadist}).

Future observations of low-redshift BLA systems with the Hubble Space Telescope/Cosmic Origins Spectrograph (COS) will improve the BLA census in several important areas.  First, five of the seven sight lines studied here (HE\,0226$-$4110, PG\,1116$+$215, PKS\,0405$-$123, PG\,0953$+$415, and PG\,1259$+$593), as well as over a dozen other AGN sight lines observed with STIS/E140M will be observed by COS as part of the Guaranteed Time Observations.  Some of the scheduled GTO observations are high-S/N spectra of BL\,Lac objects chosen specifically to search for BLAs against non-thermal power-law continua.  More AGN observations are scheduled in several large HST Cycle 17 Guest Investigator programs (PIs: Tripp, Tumlinson).  Observations of the same targets by different instruments will help sort out real BLAs from instrumental features.  Furthermore, the exquisite S/N expected from COS data, both for previously observed and new sight lines, will be crucial in determining line profiles, identifying blended systems, and finding the expected population of weak, broad systems.  Late in our analysis process, we obtained high-S/N COS/G130M observations of PKS\,0405$-$123 which shed considerable light on individual BLA candidates.  We discuss this further and show several examples of possible BLAs observed with much higher S/N in the Appendix.  While some BLA Candidates are confirmed by the COS data, there are some significant differences which suggest that the actual number of BLAss less than catalogued herein.  Finally, the high sensitivity of COS compared with STIS will allow a much larger pathlength of the low-$z$ IGM to be surveyed, increasing our statistics on intergalactic absorbers ranging from metal-line systems to BLAs.  Increases in both the \OVI\ and BLA catalogs will undoubtedly occur.  It will be very interesting to see where the WHIM baryon census stands ten years hence.


\medskip 


\noindent
We wish to acknowledge the great assistance rendered by Steve Penton in performing custom reductions of the STIS/E140M data and investigating the mysterious differences between reported reductions.  Similarly, Brian Keeney performed the nearest-galaxy searches.  Teresa Ross was instrumental in tracking down discrepancies between published line lists.  This work was supported by the COS GTO grant NNX08-AC14G from NASA, HST Archive grant AR-11773.01-A from STScI, NSF Theory grant AST07-07474, and NASA Theory grant NNX07-AG77G.

\begin{appendix}

\section{Notes on Individual Absorbers}

We present here more detailed comments on individual BLA candidates than is present in Table~2.  Each system is identified by sight line, redshift, and our BLA classification (A=Probable BLA, B=Possible BLA, C=not a BLA).  We add a fourth category here (X) for narrow \Lya\ systems which are adjacent to or blended with BLA candidates.  See Table~2 

\subsection{HE\,0223$-$4110}

\paragraph{$z=0.06083$ (B):} Strong line detected in \Lya,$\beta$ ($\Wlya=557\pm20$~m\AA, $\Wlyb=313\pm113$~m\AA) with suggestion of broad wing on blue side and possible component structure.  No metal lines to clarify component structure.  COG for full system gives $\bCOG\sim36$~\kms\ with large uncertainties.  We list this as a possible BLA since there the broad wing on the blue side suggests a broad component.  

\paragraph{$z=0.16339$ (B):} Very strong \Lya\ system.  \Lyb\ is blended with Galactic \SiII\ absorption, but appears narrow.  \Lyg\ absorption is present but weak.  COG solution doesn't converge and there are signs of multiple components.  No metal lines to illustrate component structure.

\paragraph{$z=0.20700$ (B):} Saturated, broad \Lya\ profile with possible weak, broad component seen in the wings.  \Lyb,$\gamma$ show asymetric profiles.  Strong \OVI, \CIII, \SiIII\ (DS08) and \NeVIII\ \citep{Savage05} detections as well as possible \SIV.  We adopt the \citet{Savage05} measurements for the BLA component, but list the system only as a possible BLA since the component structure is ambiguous.

\paragraph{$z=0.30930$ (B):} Strong, broad system with probable component structure in \Lya,$\beta$,$\gamma$:  $\Wlya=388\pm7$~m\AA, $\Wlyb=130\pm3$~m\AA, $\Wlyg=33\pm3$~m\AA.  Multi-valued COG solution: $\bCOG=34\pm2$~\kms\ ($\alpha,\beta$) or $\bCOG=44\pm6$~\kms\ ($\alpha,\gamma$).  Due to this ambiguity and the likely component structure, we list this as a possible BLA.  No metal lines.

\paragraph{$z=0.38420$ (C):} Strong system with ambiguous component structure in \Lya.  However, \Lyb\ profile shows two clear components.  Two or three components seen in \Lyb\ profile despite poor S/N.  COG on stronger, red, component gives $\bCOG\sim31$~\kms.  

\paragraph{$z=0.39641$ (B), $z=0.39890$ (B), $z=0.40034$ (B):}  Possible continuum ripples shown in Figure~\ref{fig_echripple} and discussed in the text.  In particular, a very similar feature appears in the data at 1702.5\AA\ in the PG\,0953$+$414 data at $z_{\rm abs}>z_{\rm AGN}$ which looks similar to the $z=0.40034$ feature toward HE\,0223$-$4110, however other spectra at higher S/N do not show this feature.

\subsection{PKS\,0405$-$123}	

\paragraph{$z=0.03196$ (B):} Broad \Lya\ profile ($\Wlya=108\pm22$~m\AA, $\bLW=57\pm8$~\kms), but weak, narrow \Lyb\ ($\Wlyb=50\pm6$~m\AA, $\bLW=35\pm4$~\kms).  COG is poorly constrained: $\bCOG>10$~\kms.  Possible BLA based on \Lya\ alone.  COS observations are consistent with the STIS/E140M data.

\paragraph{$z=0.05896$ (C):} Noisy data suggests two weak, narrow components each with $b\sim20$ \kms.  COS spectrum confirms the two components with $\bLW=23$ and 27 \kms\ are not BLAs.

\paragraph{$z=0.07218$ (C):} Very marginal feature which is not clearly \Lya.  Identified as \CIII\ $z=0.3340$ by DS08.  Absorption feature confirmed in COS data as real with $\bLW=30\pm3$ \kms.

\paragraph{$z=0.07523$ (B):} Very weak detection measured by both W06 ($\bLW=56$ \kms) and L07 ($\bLW=48\pm20$ \kms).  High-S/N COS observations show no absorption at this position ($W<11$ m\AA).

\paragraph{$z=0.08139$ (A):} Moderately strong system with $\bLW\sim53$~\kms, $\Wlya=261\pm16$~m\AA.  \Lyb\ shows $W=90\pm8$~m\AA\ but is blended with \Lye\ $z=0.18269$ ($\Wlya\sim670$~m\AA).  If this entire absorption feature is taken as \Lyb\ $z=0.08139$, then $\bCOG>22$~\kms.  However, if only part of absorption is taken as \Lyb\ ($\Wlyb\sim45\pm10$~m\AA), the COG linewidth is large but uncertain: $\bCOG=74\pm30$~\kms.  Given this ambiguity and the broad, clean \Lya\ profile, we list the absorber as a probable BLA.
 
\paragraph{$z=0.09659$ (B):} Strong \Lya\ system ($\Wlya\sim500$~m\AA) with excess in blue wing indicating possible broad component.  DS08 find $\bCOG=36\pm4$~\kms\ and W06 find $\bLW=37\pm1$~\kms\ for the entire system.  However, L07 fit a broad component to the system and we take this as a possible BLA due to the ambiguous nature of the fit.  DS08 report \OVI\ detected in both lines of the doublet associated with the system, though it is unclear to which component it may correspond.  COS observations show consistent line profile, but BLA component cannot be verified. 

\paragraph{$z=0.10298$ (C):} Very marginal feature which is not obviously a single absorber.  W06 measure $\bLW=60\pm14$~\kms.  COS data shows a clear blend of two absorbers $\bLW=27\pm5$ and $42\pm3$ \kms, thus no BLA is confirmed in this system..

\paragraph{$z=0.10419$ (C):} Very marginal detection not confirmed by DS08.  COS observations show no absorption at this position ($W<10$ m\AA).

\paragraph{$z=0.13102$ (B):} Moderate absorption feature with broad wing on red edge suggesting possible BLA component.  COS observations show absorption at this wavelength ($\bLW=35\pm1$ \kms, $\log\,N=13.18\pm0.01$), significantly narrower than that observed by STIS.

\paragraph{$z=0.13377$ (B):} Noisy data with moderately broad absorption.  COS observations show a clear absorber at this location ($\bLW=53\pm2$ \kms, $\log\,N=13.10\pm0.02$) but significantly shallower and broader than the STIS profile.

\paragraph{$z=0.13646$ (C):} Marginal line with likely components.  Single-component fit from W06 gives $\bLW=49\pm8$ \kms.  COS confirms double component structure, $\bLW=23\pm4,26\pm2$ \kms.

\paragraph{$z=0.13924$ (C):} No detection in L07.  Marginal feature in DS08 and W06.  Not clearly a single component even if real.  COS observations show two absorbers offset in velocity, but nothing corresponding to the reported line profile.

\paragraph{$z=0.15304$ (C):} Strong, broad line ($\Wlya=263\pm10$~m\AA) identified as \Lya\ by L07, but corresponding \Lyb\ nondetection ($\Wlyb<24$~m\AA) is inconsistent with a single \Lya\ absorber.  Unknown line identification.  COS confirms line profile.

\paragraph{$z=0.16121$ (C):} Strong absorber with several clear components.  Strong metal lines likely make up the flanking components leaving little room for a broad \Lya\ core.  COS confirms general profile.   Thus COS does not confirm this BLA.

\paragraph{$z=0.16678$ (B) and $z=0.16714$ (C):} Strong \Lya\ complex ($W_{\rm Ly\alpha}\sim650$~m\AA) with an asymmetry on the blue wing.  Line can be fit with broad a broad and narrow component, but decomposition is suspect.  \Lyb\ ($W_{\rm Ly\beta}\sim450$~m\AA) shows no sign of the broad, blue component.  Strong system ($z=0.16714$) shows obvious component structure in various metal lines.  The system shows strong \OVI\ and \NV\ absorption (DS08), but it is unclear with which component it is associated.  COS observations show a consistent line profile, but cannot confirm or deny the presence of a BLA component.
 
\paragraph{$z=0.17876$ (B):} Asymmetric \Lya\ line with very marginal \Lyb\ detection.  COG analysis by DS08 gives $\bCOG=18^{+44}_{-7}$~\kms.  Apparent Optical Depth (AOD) line fits to \Lya\ line gives much broader profile ($\bLW\sim60$~\kms; W06 measures $\bLW=58\pm6$~\kms).  Well fit by an offset broad+narrow pair ($\bLW=57$, 14~\kms), but inconclusive.  We take the AOD measurements as concensus in this case.  COS observations confirm the general profile shape of this absorber and consistent line measurements.

\paragraph{$z=0.18269$ (B):} Strong \HI\ system ($\Wlya\sim690$~m\AA) with at least two components, possibly more.  \Lyb\ is blended with Galactic \Lya.  \Lyg\ shows strong, moderately asymmetric profile ($\Wlyg=139\pm11$~m\AA).  COG gives $\bCOG=49^{+11}_{-7}$~\kms, but probable component structure relegates this to a possible BLA.  System shows several \OVI\ components.  COS observations confirm the line profile of this strong absorber, but cannot confirm or deny the presence of a BLA.

\paragraph{$z=0.19086$ (B):} Weak, noisy absorber with possible component structure.  COS observations do not show an absorption line at this position ($W_r<22$ m\AA).

The COS ERO spectral coverage ($1132<\lambda<1468$) ends at this maximum redshift.

\paragraph{$z=0.24513$ (B) and $z=0.24553$ (X):} Pair of well-separated \Lya\ absorbers.  System at $z=0.24513$ appears narrow ($\bLW=30\pm5$~\kms) but both L07 and W06 measure $b\approx55$~\kms.  We adopt a consensus $b$ and $N$ intermediate between the two measurements to account for possible continuum and reduction differences.

\paragraph{$z=0.32500$ (C):} Marginal detection, however there is a correlation between features in \Lya,$\beta$.  Probably multiple components.  DS08 report $\bCOG=21$~\kms, however we find this to be more reasonable as a lower limit and the COG solution doesn't fit the absorption profile very well.  W06 report $\bLW=81\pm11$~\kms.  Higher S/N data on this absorber should prove interesting.  Given the marginal nature of this feature and the likely component structure, we list it as a non-BLA.

\paragraph{$z=0.35092$ (B), $z=0.35149$ (X):} Strong \Lya\ system flanked by \SiIII\ $z=0.36079$ and \Lya\ $z=0.35149$.  \Lya,$\beta$ curve of growth gives $\bCOG=56\pm15$~\kms\ (DS08), but \Lya,$\gamma$ curve of growth gives $\bCOG=27\pm5$ and seems to better match the profile.  W06 and L07 measure $\bLya=38\pm2$~\kms\ and $\bLya=40\pm8$~\kms, respectively.  Both \Lya\ and \Lyb\ show asymmetric profiles suggesting multiple components.  We accept the \Lya\ measurements as consensus in this case and list the system as a possible BLA.

\paragraph{$z=0.36150$ (C) and $z=0.36079$ (X):} Strong, blended \Lya\ systems with an ambiguous component structure and noisy data.  Two-component fit gives $\bLW=75\pm10$, $39\pm5$~\kms\ for the $z=0.36150$ and $z=0.36079$ components, respectively.  The $z=0.36150$ component, however, doesn't appear in \Lyb\ ($\Wlyb\le9$~m\AA); based on the \Lya\ component, the expected \Lyb\ line should be $\Wlyb\sim50$~m\AA.  Meanwhile, a \Lyb,$\gamma$ curve of growth fit to the the $z=0.36079$ component (ignoring the noisy, possibly blended \Lya\ profile) gives $\bCOG=25\pm3$~\kms, $\log\,N=15.18\pm0.13$~\cd.  We conclude that neither system is a BLA.

\paragraph{$z=0.40886$ (B):} Strong \Lya\ system ($\Wlya=430\pm10$~m\AA) at the long-wavelength, noisy end of the STIS/E140M range.  \Lyb,$\gamma$ also present ($\Wlyb=126\pm15$~m\AA, $\Wlyg=68\pm10$ m\AA), both with asymmetric profiles.  \Lya,$\beta$ and \Lya,$\gamma$ COG solutions inconsistent but generally $\bCOG\sim40$~\kms.  Possible BLA, but probable component structure.

\subsection{HS\,0624$+$6907}

\paragraph{$z=0.04116$ (C):} Weak feature consistent with \SiIII\ $z=0.06346$, a system in which many other ionic lines are also detected.

\paragraph{$z=0.05437$ (B), $z=0.05515$ (B), and $z=0.05484$ (C):} Strong \Lya\ system ($\Wlya\sim450$~m\AA) with several plausible BLA subcomponents.  The $z=0.05437$ line appears as a very marginal, noisy absorption on the blue wing while the $z=0.05515$ component is an excess on the red side of the main absorption line.  The strong central component at $z=0.05484$ is fit well by a $\bLW\sim35$~\kms\ component despite the $\bCOG=45^{+25}_{-9}$~\kms\ value reported in DS08.  The narrow feature at 1283.15 \AA\ can be unambiguously identified as \SiIII\ $z=0.0635$.  The BLA nature of these lines relies on the details of the component structure which is quite ambiguous and difficult to disentangle. 

\paragraph{$z=0.06346$ (C):} Strong \Lya\ system with measured $\bLW>40$~\kms\ but $\bCOG<40$~\kms\ with \OVI\ detected in both lines of the doublet (DS08).  \HI\ profile looks symetric, but there are two clear components in \SiIV, \CIV, \SiIII, and \SIII.
 
\paragraph{$z=0.13597$ (B):} Weak double profile.  Feature at 1381.4\AA\ is \Lyb\ $z=0.3468$ but broader feature at 1381.0\AA\ is a plausible BLA with low-significance \OVI\ detections at both 1032 and 1038\AA.  Possible BLA based on the measured line though the large error bars and \OVI\ detection could argue for this being a Probable BLA.

\paragraph{$z=0.30994$ (B):} Broad, weak absorption feature identified as \Lya\ by both L07 and DS08.  However considerable continuum uncertainty yields highly incompatible $b,N$ solutions.  Possible, low-significance \OVI\ detected as well.  In any case, profile is asymmetrical and not obviously a single component.

\paragraph{$z=0.31045$ (C), $z=0.31088$ (C):} Broad, asymmetric features which are not obviously real.  Possible continuum differences.  

\paragraph{$z=0.31790$ (B):} Moderate, narrow \Lya\ line with asymmetry on the red wing and good \OVI\ detections in both lines of the doublet.  Possible multiphase system and/or continuum fit uncertainties.  Possible BLA.

\paragraph{$z=0.32089$ (C):} Strong, moderately-narrow line is a poor fit to DS08 COG solution ($\bCOG=44^{+13}_{-14}$~\kms).  \Lya\ absorption is anomalosly large given observed \Lyb,$\gamma$, however there are no other obvious contributors to the \Lya\ line strength.  We confirm the $\bLW=31\pm1$~\kms\ solution of L07 for this absorber.

\paragraph{$z=0.33976$ (A):} Strong, broad system with no obvious components and detections in \Lya,$\beta$,$\gamma$.  Good COG solution gives $\bCOG=41\pm6$~\kms.  \OVI\ detected in both lines of the doublet (DS08).

\subsection{PG\,0953$+$415}

\paragraph{$z=0.04382$ (C):} Reported as \OVI\ \lam1032 at $z=0.22974$ by \citet{Tripp08}.  Interestingly, this is a ``blind'' \OVI\ system with little or no accompanying \Lya\ absorption ($\Wlya<7$~m\AA).
 
\paragraph{$z=0.05879$ (B):} Narrow line with a possible excess in line wings suggesting a multiphase system.  Very ambiguous decomposition.  We take the L07 values as the consensus. 

\paragraph{$z=0.12784$ (C):} Claimed broad blue wing on a narrower \Lya\ system.  We see no good evidence for this system in our reduction of the data.

\paragraph{$z=0.17985$ (B):} Weak single or double absorption.  Potentially fit as single BLA or two narrow components ($\bLW\sim20$).

\paragraph{$z=0.19126$ (B):} Asymmetric \Lya\ profile with uncertain decomposition.  Formal AOD linewidth is $\bLW\approx66$~\kms\, however, two components ($\bLW=40$, 32~\kms) provides a better fit.  We adopt $\bLW=40:$~\kms\ and $\log\,N=13.3$ \cd\ as consensus values.

\paragraph{$z=0.19361$ (C):} Strong narrow \Lya\ system with very weak \Lyb,$\gamma$ counterparts.  DS08 report a \Lya,$\beta$ solution ($\bCOG=24^{+4}_{-3}$~\kms, $\log\,N=14.15\pm0.04$ \cd) which is stronger than the data can easily support.  We measure $\bLW=37\pm2$~\kms\ instead and adopt that value here.

\subsection{PG\,1116$+$215}

\paragraph{$z=0.08587$ (B), $z=0.08632$} Pair of very shallow, ripples in the continuum with $\bLW\sim55$ and $\bLW\sim34$~\kms, respectively, both with considerable uncertainty.  Broader feature is possible BLA.

\paragraph{$z=0.09279$ (A):} Very weak, broad system blended with two weak, narrow features (possible Galactic \ion{C}{1}).  While profile looks questionable, it is more obvious when looking at a broader wavelength range.  Eliminating the two sharp features gives a very plausible BLA.  

\paragraph{$z=0.13370$ (A):} BLA with a low-significance \OVI\ \lam1032 detection ($\log\,N_{\rm OVI}\approx13.2\pm0.3$ \cd).  Potentially fit as two components ($\lambda\approx1378.3,1378.7$) but $\bLW\sim50$~\kms\ even in this case.

\subsection{PG\,1259$+$593}

\paragraph{$z=0.00229$ (A):} \Lya\ line appears on the edge of the broad Galactic \Lya\ trough and \Lyb\ is blended with an \OI\ airglow line in \FUSE\ data.  \citet{Richter04} used night-only data to get a good \Lyb\ measurement and constrains $\bCOG=44^{+9}_{-4}$~\kms\ which we adopt here. 

\paragraph{$z=0.04606$ (C):} Very strong system with obvious components in \OVI, \CIV, \SiIII, \SiIV.  Richter et al. (2004) report two \HI\ components, both with $\bCOG<40$. 

\paragraph{$z=0.14852$ (A):} Strong nearly triangular line profile with detections in \Lya,$\beta$.  Curve of growth consistent with BLA $\bCOG\approx57$~\kms\ though with large errors.  $\blyb\sim36$~\kms.

\paragraph{$z=0.19573$ (B), $z=0.19620$ (X):} Double absorption profile.  Shallow, blue feature at 1453.6 \AA\ is plausible BLA candidate ($\bLW=46\pm19$~\kms) and was reported as a component to the $z=0.1962$ system by \citet{Richter04}.  Marginal \NV\ detection.  

\paragraph{$z=0.21136$ (C) and $z=0.43569$}  Richter et al. (2004) identify the absorption at 1472.6 \AA\ as \Lyb\ $z=0.43569$ and high-res FOS spectra confirm \Lya\ line at this redshift.  Absorption at 1242.5 \AA\ where \Lyb\ $z=0.21136$ would be expected is blended with \Lya\ $z=0.02217$ and unrecoverable.  Can't confirm.

\subsection{H\,1821$+$643}

\paragraph{$z=0.11133$ (A), $z=0.11167$ (X):} Double absorber profile with one broad, one narrow system ($\bLW=52,31$~\kms).  L07 measure this as a single system ($\bLW=88\pm14$~\kms).  One BLA likely present.

\paragraph{$z=0.12147$ (A), $z=0.12117$ (X):} Strong and highly asymmetric absorber.  Red wing \Lyb\ is hinted at in the data, but at low significance.  System shows $\bCOG>40$~\kms, but there are clearly multiple components.  Weak \OVI\ absorption is reported in the system \citep{Tripp01,Tripp08,Oegerle00} aligned with broad \Lya\ wing, however, DS08 do not confirm this detection.

\paragraph{$z=0.14754$ (B) and $z=0.14776$ (X):} Blended system of two, easily separable components ($\bLW\approx42,20$~\kms).  Marginal \Lyb\ detection for both components but COG solution is very poorly constrained ($\bCOG=22^{+\infty}_{-12}$~\kms) for the $z=0.14754$ system.

\paragraph{$z=0.21326$ (A):} Strong system with \Lya,$\beta$ detections ($\bCOG=42^{+5}_{-4}$~\kms) though with some hint of component structure.  Strong \OVI\ detection in both lines of the doublet.  The entire absorber may not represent hot gas, but there is probably a BLA component.

\paragraph{$z=0.22480$ (C):} Very strong system ($\Wlya\sim800$~m\AA) with clear components in higher Lyman lines.  \OVI\ detected.  Not confirmed by L07, but clearly not a BLA.  

\paragraph{$z=0.22616$ (B):} Moderate absorption feature ($\Wlya=150\pm8$~m\AA) identified as \Lya\ by L07.  However, corresponding \Lyb\ non-detection shows $\Wlyb\le10$~m\AA, inconsistent with a single \Lya\ system.  Possible \OVI\ absorption ($\log\,N_{\rm OVI}=13.4\pm0.1$ \cd) in both lines of the doublet offset $\sim+60$~\kms\ from the purported \HI\ system.  Probable multiple components or misidentification.

\paragraph{$z=0.25814$ (B):} Strongly asymmetric profile suggests a multiphase \HI\ system.  Profile is reasonably fit by two components: $\bLW\approx56$, 17~\kms.  Total line shows $\Wlya=141\pm3$~m\AA\ and corresponding weak \Lyb\ absorption ($\Wlyb\approx24$~m\AA) gives a curve-of-growth solution $\bCOG=30\pm7$~\kms\ for the combined system.  In light of this, the component fits seem reasonable and we list this as a possible BLA.  

\paragraph{$z=0.26658$ (A):} Strong system with \OVI\ detected in both lines of the doublet.  Poorly-constrained COG gives $\bCOG=68\pm30$~\kms.  Individual lines show $\bLW=46$~\kms\ and $\bLW\sim40$~\kms.  Probable BLA despite poorly-constrained COG.

\end{appendix}

\end{document}